\documentclass[journal]{IEEEtran}
\ifCLASSINFOpdf
  \usepackage[pdftex]{graphicx}
\else
\fi
\usepackage{cite}
\usepackage{amsmath}
\usepackage{amsfonts} 

\usepackage{algorithmic}
\usepackage[vlined,linesnumbered,ruled]{algorithm2e}
\usepackage{array}
\usepackage{color}
\usepackage{subfigure}
\usepackage{mdwtab}
\usepackage{booktabs}
\usepackage{multirow}
\usepackage{hyperref}
\usepackage[hyphenbreaks]{breakurl}
\usepackage{threeparttable}
\usepackage[ruled,linesnumbered]{algorithm2e}
\usepackage{rotating,setspace}
\usepackage{graphics,epsfig}
\usepackage{epstopdf}
\usepackage[utf8]{inputenc}
\usepackage[english]{babel}

\usepackage{amsthm}
\usepackage{amssymb}
\newcommand*{\QEDA}{\null\nobreak\hfill\ensuremath{\blacksquare}}

\usepackage{mathtools, nccmath}
\DeclarePairedDelimiter{\nint}\lfloor\rceil

\usepackage{physics}

\begin{document}
\title{Enhanced Time-Frequency Representation \\and Mode Decomposition}

\author{Haijian~Zhang and Guang~Hua
        \vspace{-1.\baselineskip}
\thanks{H. Zhang and G. Hua are with the School of Electronic Information, Wuhan University, Wuhan 430072, China.}
}

\markboth{Time-frequency analysis and its applications}{}
\maketitle

\begin{abstract}

Time-frequency representation (TFR) allowing for mode reconstruction plays a significant role in interpreting and analyzing the nonstationary signal constituted of various modes. However, it is difficult for most previous methods to handle signal modes with closely-spaced or spectrally-overlapped instantaneous frequencies (IFs) especially in adverse environments. To address this issue, we propose an enhanced TFR and mode decomposition (ETFR-MD) method, which is particularly adapted to represent and decompose  multi-mode signals with close or crossing IFs under low signal-to-noise ratio (SNR) conditions. The emphasis of the proposed ETFR-MD is placed on accurate IF and instantaneous amplitude (IA) extraction of each signal mode based on short-time Fourier transform (STFT). First, we design an initial IF estimation method specifically for the cases involving crossing IFs. Further, a low-complexity mode enhancement scheme is proposed so that  enhanced IFs better fit underlying IF laws. Finally, the IA extraction from signal's STFT coefficients combined with the enhanced IFs enables us to reconstruct each signal mode. In addition, we derive mathematical expressions that reveal optimal window lengths and separation interference of our method. The proposed ETFR-MD is compatible with previous related methods, thus can be regarded as a step toward a more general time-frequency representation and decomposition method. Experimental results confirm the superior performance of the ETFR-MD when compared to a state-of-the-art benchmark. 

\end{abstract}


\IEEEpeerreviewmaketitle

\section{Introduction}

\IEEEPARstart{T}{ime}-frequency representation (TFR) of nonstationary signals constituting of different modes has received considerable research attention. More importantly, separation of multi-mode signals, often referred to as mode decomposition, has also been studied for several decades. In practical applications, it is desirable to decompose the signal of interest into its constituent modes, each pertaining to an intrinsic component. Therefore, many signal processing challenges boil down to the problem of signal representation and decomposition, which is still under investigation as well in methodological aspects as in concrete applications. The focus of this paper is put on representing and decomposing multi-mode frequency-modulated (FM) signals with closely-spaced or spectrally-overlapped instantaneous frequencies (IFs), which is a difficult problem yet to be adequately resolved.

A wealth of information could be found in the time-frequency (TF) analysis literature \cite{2013LjubisaStankovic,2016ivBoualem,2018CambridgePatrick,SEJDIC2009153}, such as linear and quadratic TFRs \cite{Hlawatsch127284}, polynomial TFRs \cite{boashash1998polynomial,4034118Bi}, etc. Most of recent  studies developed so far are low-order TFRs, mainly including linear short-time Fourier transform (STFT), linear fractional Fourier transform (FRFT) \cite{Liu6942239,Shi9091935}, quadratic pseudo Wigner-Ville distribution (PWVD) and  classical Cohen's class TFRs \cite{30749Cohen}. Despite high TF resolution of nonlinear TFRs, they generally suffer from  cross-term (CT) interference and irreversible TF transform, which cause a serious barrier to mode reconstruction. Prior studies that have noted the importance of mode deconstruction emphasize on linear TFRs due to their low computation cost and invertibility  \cite{Griffin1164317}. 

The theory of linear TFRs has been investigated quite intensively during the past decades. Wavelet transform (WT) \cite{daubechies1990wavelet}, which is regarded as a classical linear TFR, employs adaptive window sizes to improve TF resolution compared to linear STFT. Signal representation was afterwards improved by implementing post-processing based on WT or STFT. The reassignment (RS) methods  \cite{382394,Auger6633061} assign the average energy in certain domains to the gravity center of energy distributions to gain IF trajectories. However, a major problem associated with RS related methods is the infeasibility of mode reconstruction. Further, the synchrosqueezing transform (SST) \cite{Oberlin6853609} derived on WT \cite{DAUBECHIES2011243} or on STFT \cite{Thakur100798818} squeezes TF coefficients into the IF trajectory only in  frequency direction instead of in both time and frequency directions for RS, with the aim of reconstructing signal modes. In recent years, advantages to mode representation and reconstruction have resulted in an increasing interest in the development of SST-based methods \cite{Wang6574272,oberlin2015second,Pham7885114,Yu8458385}.
Inspired by the SST, Yu \textit{et al.} proposed the synchroextracting transform (SET) \cite{yu2017synchroextracting}, which employs a newly developed  synchroextracting technique to generate a more energy concentrated TFR, and meanwhile the invertible SET allows for mode reconstruction. Thanks to the simplicity and immunity to CTs of linear TFRs, Abdoush \textit{et al.} \cite{abdoush2019adaptive} developed two linear TF transforms, based on which an adaptive IF estimation method for noisy multi-mode signals was implemented.

\begin{figure*}[!t]
\hspace{-3mm}
\subfigure[Five closely-spaced FM modes.]{
\includegraphics[width=2.5 in]{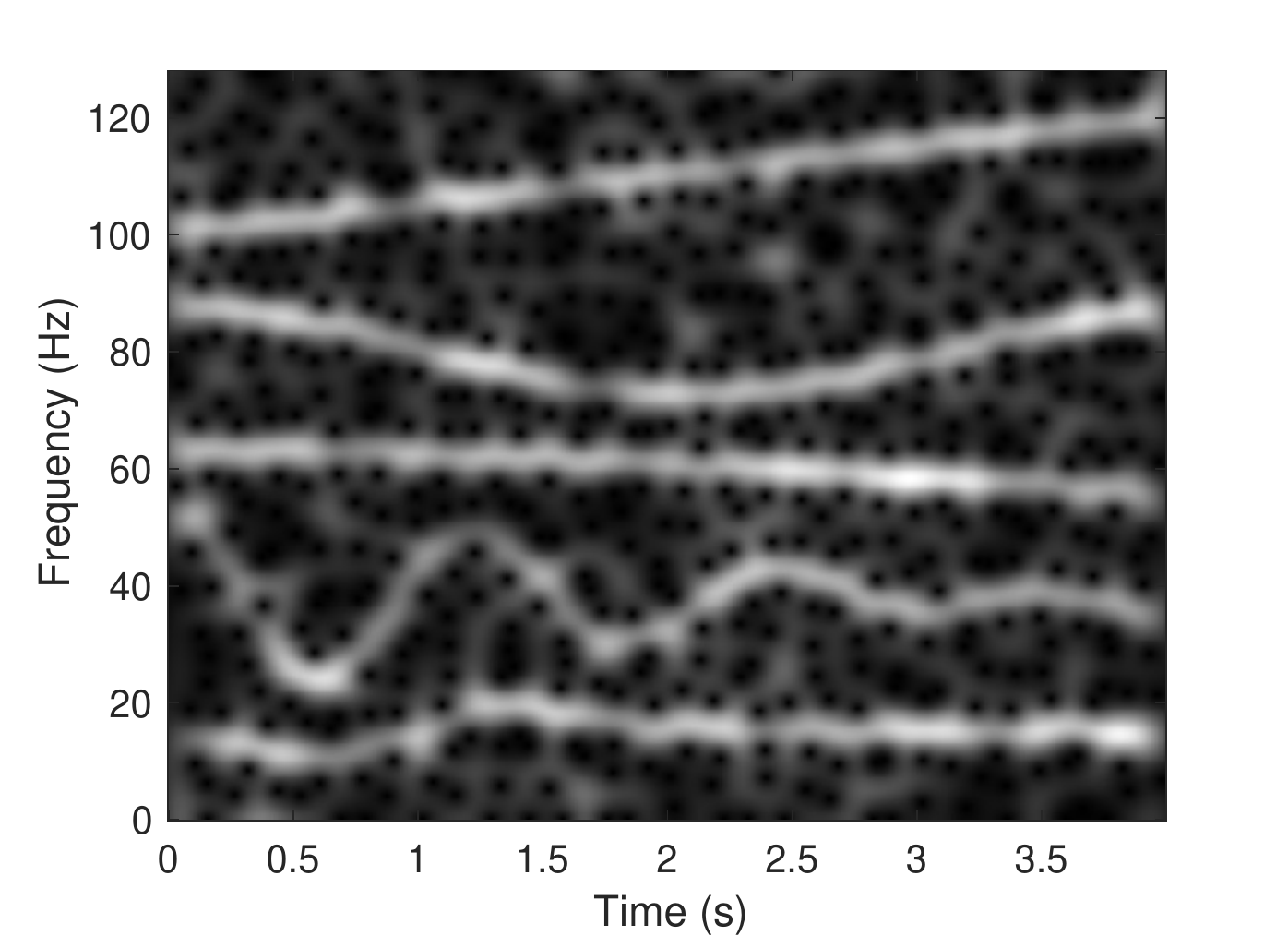}
}
\hspace{-7mm}
\subfigure[Two crossing weakly-modulated FM modes.]{
\includegraphics[width=2.5 in]{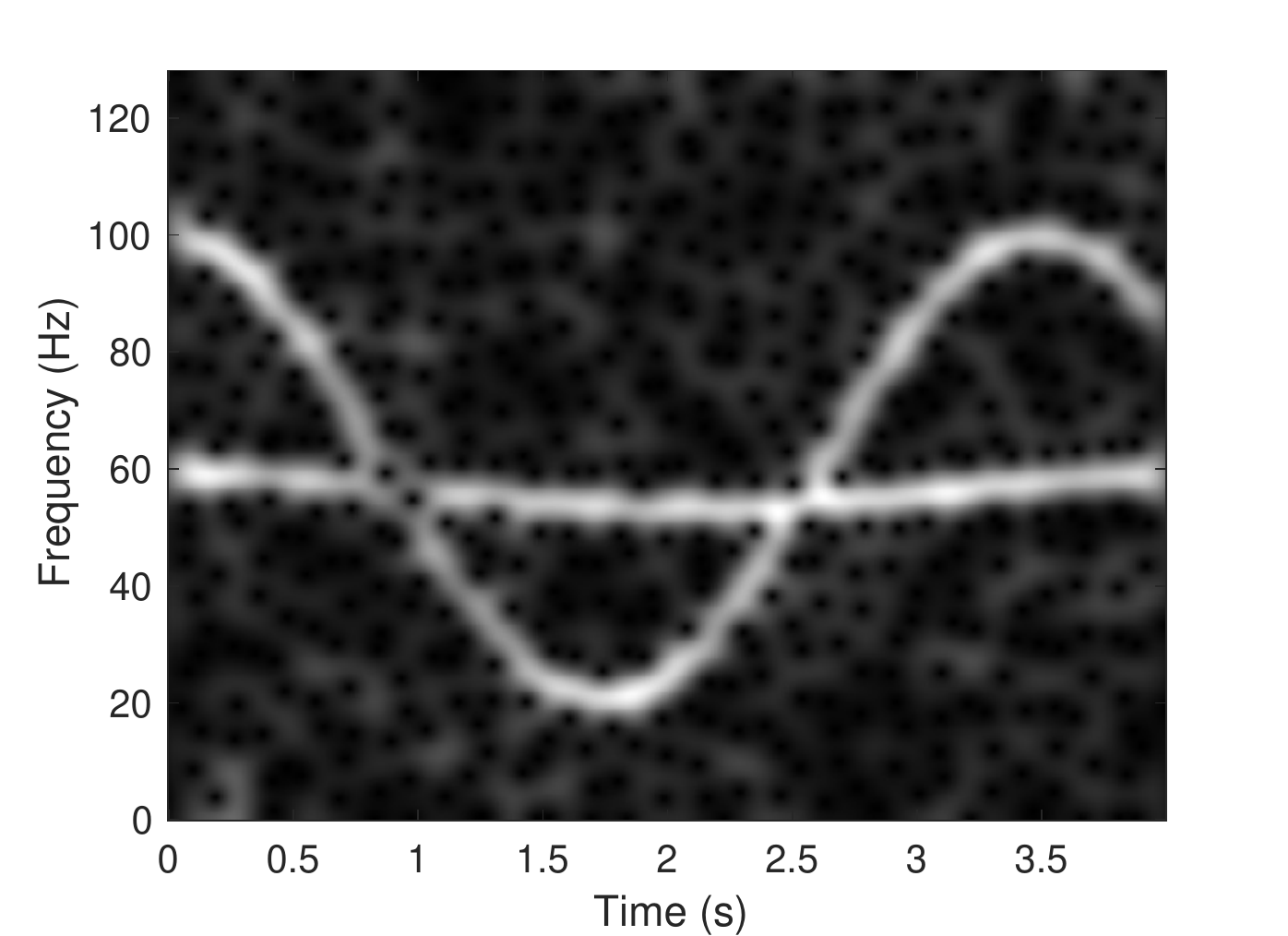}
}
\hspace{-7mm}
\subfigure[Two crossing highly-modulated FM modes.]{
\includegraphics[width=2.5 in]{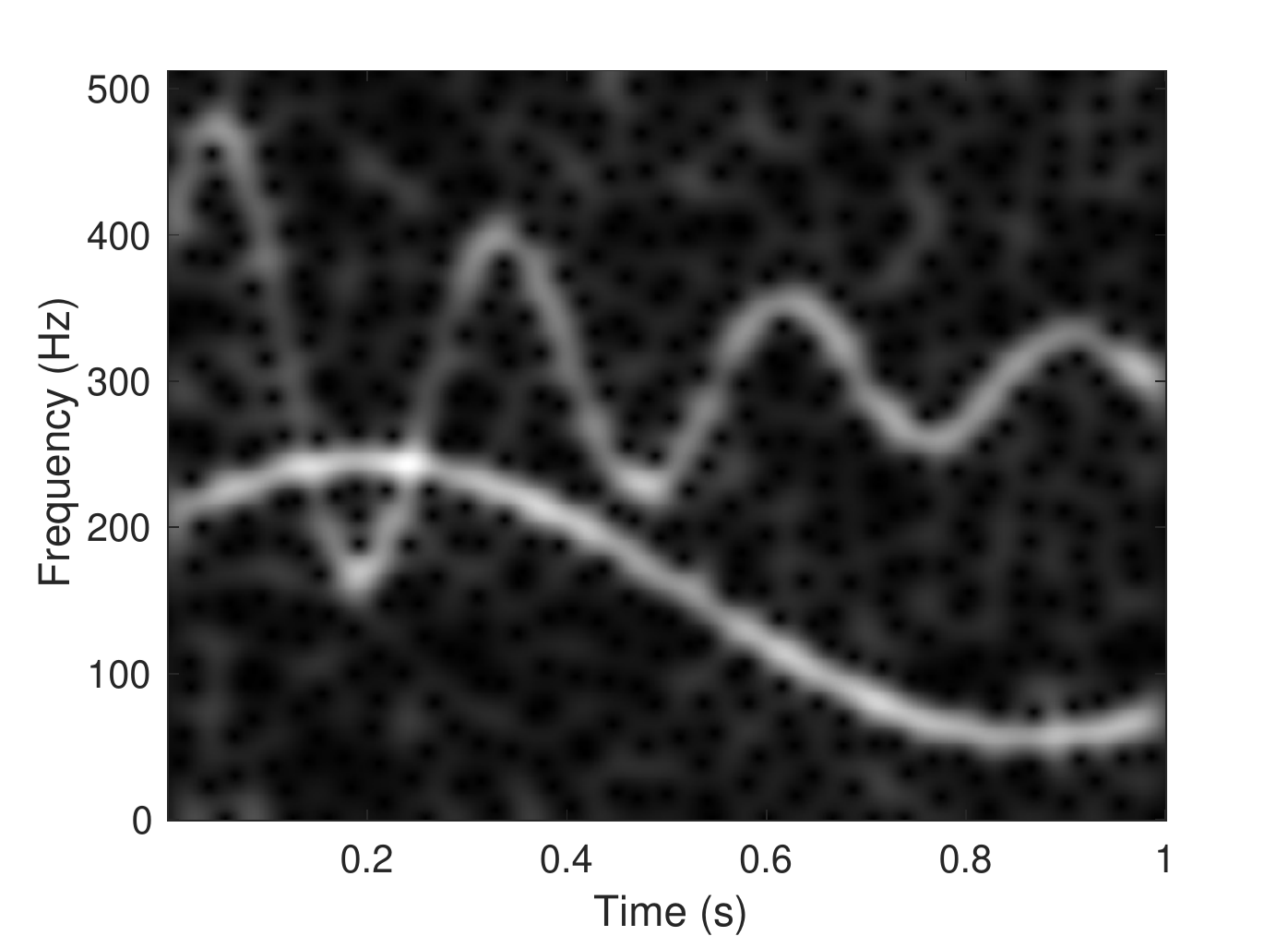}
}
\caption{STFTs of three typical multi-mode signals at SNR = 5 dB: \textbf{(a)} Five closely-spaced FM modes including one linear FM (LFM), two sinusoidal FM (SFM), one damped SFM, and one nonlinear FM. \textbf{(b)} Two spectrally-overlapped and weakly-modulated SFM modes. \textbf{(c)} Two spectrally-overlapped and strongly-modulated SFM modes.}
\label{fig1}
\end{figure*}

Although the above studies work remarkably well, they have to rely on certain separation condition, indicating that  the multi-mode signals to be addressed should have non-overlapped and well separated IFs. This requirement leads to the problem of how to deal with multi-mode signals with spatially-adjacent or crossing IF laws especially when the signal-to-noise ratio (SNR) level is exceedingly low, e.g., the illustrative three signals in Fig. \ref{fig1}. 
Though application of TFRs in the filed of mode decomposition has proliferated in recent years, effort in analyzing complicated signals in strong noise environment is lacking. To implement mode decomposition, accurate IF estimation of each mode is thought of as the key factor. Diverse IF estimation methods have been well explored and developed. The pioneer work of IF estimation can be traced to the works for mono-mode signals \cite{Boashash135378}, and then developed for non-overlapped multi-mode signals \cite{Carmona640725,Carmona740131,DJUROVIC2004631,Hussain1018780,IATSENKO2016290,zhu2020adaptive} and overlapped multi-mode signals \cite{RANKINE20071234,7073495Zhang,KHAN2020107728,ZHU2020102783}. While existing IF estimation methods have achieved a certain effect, it is not a trivial task to match their ability for complicated signals with crossing IFs. Thus, more works should consider the extension of a general IF estimator to deal with spectrally-overlapped multi-mode signals. 

In this paper, we propose an enhanced TFR and mode decomposition (ETFR-MD) method, with the objective of addressing the multi-mode signals consisting of multiple closely-spaced or spectrally-overlapped modes. A major concern of the proposed ETFR-MD is to respectively reconstruct IF and instantaneous amplitude (IA) of each mode.  To achieve this, we firstly design an initial multi-mode IF estimation method based on STFT instead of on SST-based TFRs \cite{DJUROVIC1999115,IATSENKO20151,zhu2020adaptive}. To compensate  fluctuating or unsmooth IF estimates, an iterative filtering method is then developed to improve IF estimation accuracy \cite{ZHANG2015141}. On the other hand, the IA of each mode is extracted from  STFT coefficients with a Gaussian window. The combination of enhanced IFs and extracted IAs accomplishes the representation and decomposition of all signal modes. Finally, the proposed ETFR-MD method is verified and justified through experimental study. 

The structure of this paper is organized as follows. In Section II, we introduce the signal model and general problem. This is followed by a description of the proposed ETFR-MD, and a detailed presentation of how IF and IA of each mode are recovered in the STFT framework is provided in Section III. We then derive in Section IV the expressions of optimal window lengths and conduct interference analysis, which are necessary for the understanding of our method. Section V presents the experimental results with the emphasis on multi-mode signals consisting of close or crossing IFs. Finally, the ETFR-MD method is summarized in Section VI.

\section{Signal Model and General Problem}

Without loss of generality, we consider noisy multi-mode signals with $I$ time-varying modes:
\begin{align} \label{TSP_eq1}
s[n]=& \sum_{i=1}^I x_i[n]+\epsilon[n]
\notag\\
=& \sum_{i=1}^I \rho_i[n]\cos\bigg( 2\pi T \sum_{m=0}^{n}f_{i}[m] +\theta_i\bigg)+\epsilon[n],
\end{align}
where $I$ is the number of modes, $\rho_i[n]$  is the  IA of the $i$-th mode, and $T$ denotes sampling interval. The IF of the $i$-th mode, $f_{i}[n]$, is defined by the derivative of the phase of its corresponding analytic signal. 
The random phase $\theta_i$ corresponding to the $i$-th mode is uniformly distributed on $[0,2\pi)$, and  $\epsilon[n]$ is the zero-mean additive white noise. The signal model in (\ref{TSP_eq1}) is quite general  to accommodate a variety of practical signals.

Real-world signals are often received with arbitrary number and type of modes, indicating that the IF curves might be closely distributed or even intersecting. Instead of giving a narrowly defined model which does not fit well with practical scenarios, we attempt to relax requirement of the signal defined in (\ref{TSP_eq1}). For the sake of concentrating on dealing with the complicated cases in Fig. \ref{fig1}, the signal to be handled is assumed to meet the following assumptions: \textbf{i)} the number of signal modes is unknown; \textbf{ii)} the signal might be contaminated by strong noise; \textbf{iii)}  IF laws of different modes could be arbitrarily distributed, i.e., sometimes IF intersection occurs.  The goal of the proposed ETFR-MD method is to respectively enhance and recover IF and IA information of each mode, i.e., $\rho_i[n]$ and $f_{i}[n]$, $i=1,\ldots,I$.

\section{Enhanced Time-Frequency Representation and Mode Decomposition}

\begin{figure}[!t]
\hspace{-2mm}
\includegraphics[width=3.5 in]{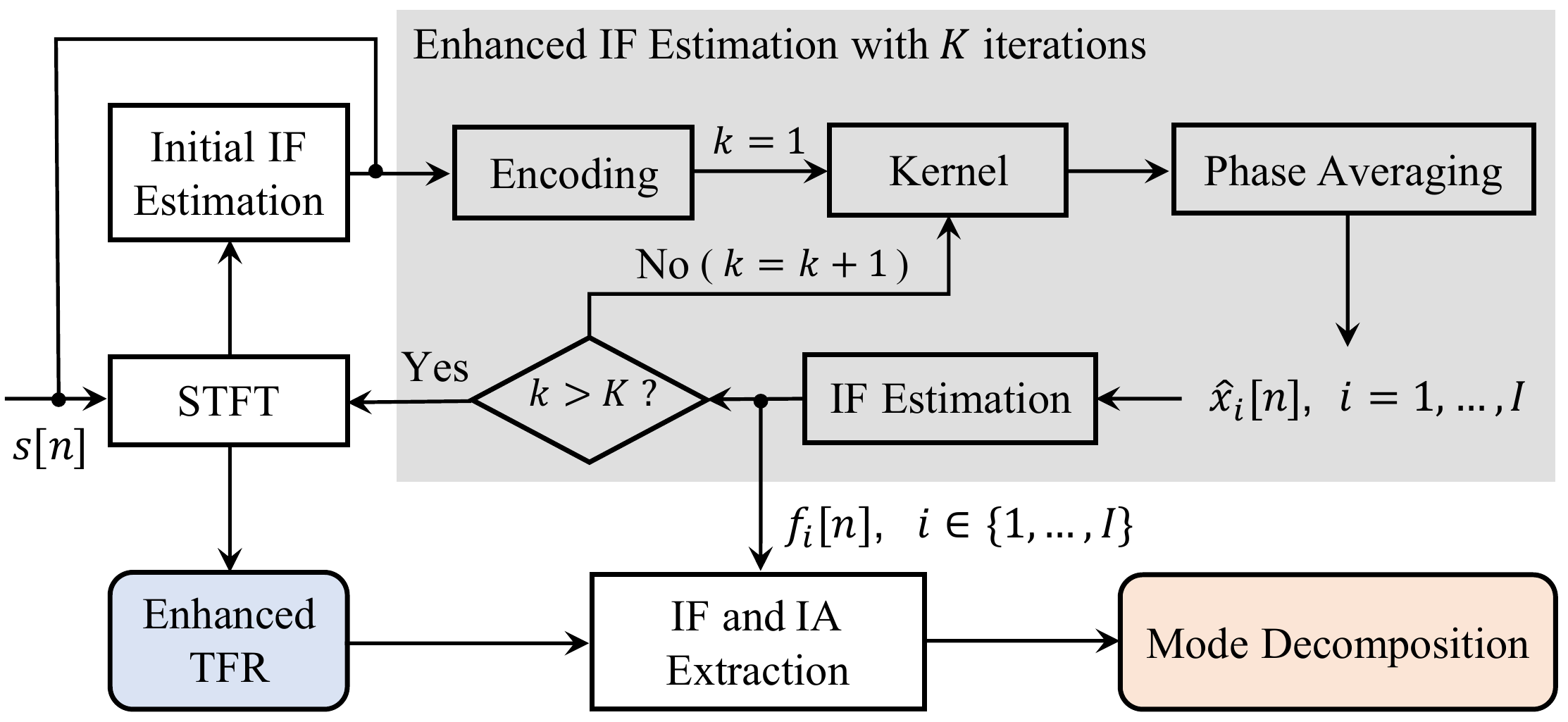}
\caption{Schematic diagram of the enhanced time-frequency representation and mode decomposition (ETFR-MD).}
\label{fig2}
\end{figure}

The basic process of the proposed ETFR-MD method is depicted in Fig. \ref{fig2}, wherein the noisy signal $s[n]$ defined in (\ref{TSP_eq1}) serves as the input, and the output of ETFR-MD  could be either enhanced TFR or decomposed signal modes (in highlighted round boxes). Note that our emphasis is placed on the enhancement of  IF estimation (the gray area in the figure), by which the fluctuating property of initial IF estimate can be mitigated. An additional benefit of  IF enhancement is that the IF information of each mode can be individually recovered with almost negligible interference from other modes. Afterwards, the IA information of each mode is extracted from Gaussian window based STFT, which offers a simple solution to the IA recovery.

In the following subsections, we first introduce an effective IF estimation method to solve complex situations involving close and crossing IFs. Second, the initial IF estimation is  refined by implementing an iterative filtering method, so that the IFs of all signal modes are enhanced. Last, the enhanced IFs and the extracted IAs derived from STFT are combined for the purpose of mode representation and decomposition.

\subsection{Initial IF Estimation}

In the literature, few works have been reported to deal with multi-mode signals with overlapped IFs. To solve this problem, we extend the work of Djurovi\'{c} and Stankovi\'{c}   \cite{DJUROVIC2004631}  originally designed for non-overlapped signal modes, and propose a smooth-and-curvature constraint based IF estimation method for both non-overlapping and overlapping cases, aiming at finding a nonparametric STFT-based solution without requiring prior knowledge of signal's IF laws. To be specific, three penalty functions are considered, and Fig. \ref{fig3} exhibits an illustrative example to explain the three IF constraints in the presence of crossing IFs and low SNR.

\subsubsection{Penalty for IF Energy}
Current IF estimation methods can be classified as phase differencing, signal modeling, phase modeling and TFR based methods, which can be also categorized into parametric and nonparametric ones. Parametric methods are complicated and time-consuming in the case of distinct IF laws, furthermore, they are only suitable for IF estimation   when the signal model is known. In most situations, the IF exhibits unforeseen variations and the parametric model is unknown. Recently, some nonparametric methods were proposed, among which a common way is to track  IF curves by detecting the maxima positions of all time-slices in STFT plane, which has been demonstrated to be useful for IF tracking.

\subsubsection{Penalty for IF Continuity}
In the case of low SNRs, the fluctuation in IF estimates becomes significant, and spurious estimates (outliers) far from true IFs frequently occur. Therefore, the continuous property  is incorporated into the IF detection to punish improbable outliers and generate continuous frequency estimates. We combine the TF energy maxima detection with minimization of IF variations between consecutive TF samples, and the basic idea comes from the problem of edge-following in digital image processing. The main criteria are:  IF should pass through as many as possible points of STFT with the highest magnitudes, while IF variation between two consecutive points should not be too fast. To relieve computational burden, the IF estimation method can be realized recursively as an instance of the Viterbi algorithm. 

\subsubsection{Penalty for IF Ridge Curvature}
\begin{figure}[!t]
\centering
\includegraphics[width=2.8 in,height=1.8 in]{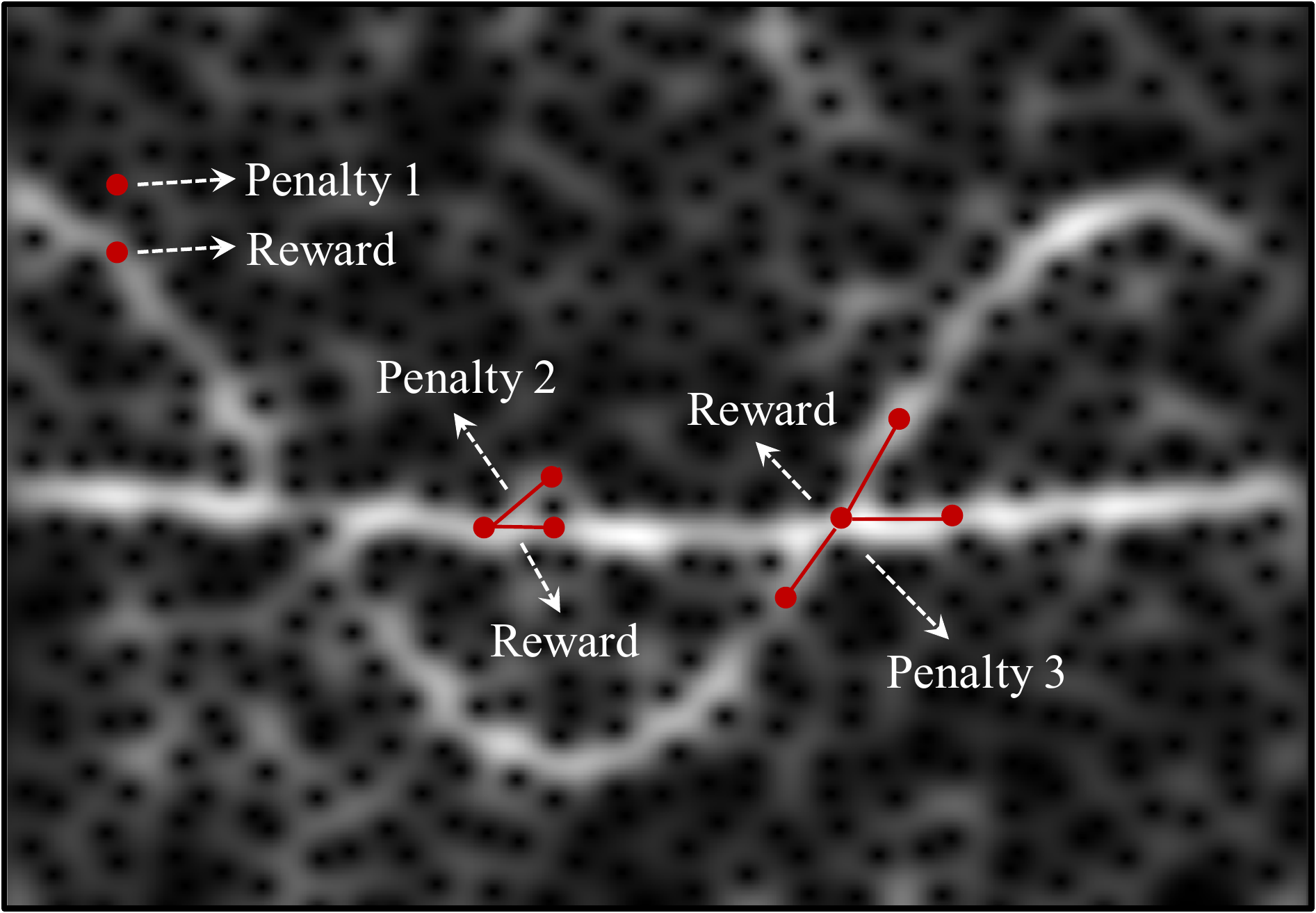}
\caption{Illustration of the penalty functions for initial IF estimation of two spectrally-crossing sinusoidal FM modes in STFT domain (SNR = 0 dB).}
\label{fig3}
\end{figure}

Nevertheless, only considering the energy and continuity constraints will not work well for multi-mode signals with IF intersections. To tackle this issue, our method not only considers continuous property between consecutive samples, but also minimizes the effect of large curvature of estimated IF segments (e.g., \textit{penalty 3} in Fig. \ref{fig3}) by looking at the overall IF trajectory structure without prior knowledge of  IF laws. This extended method enables a smoothed IF detection for multi-mode signals with non-overlapping or overlapping IF laws, and is also valid for mono-mode signals.

According to the above analysis, the essential idea of our method is to find an optimal path from the beginning time-slice to the ended time-slice in STFT plane. This optimal path corresponds to the expected IF estimate which minimizes the following expression:
\begin{align} \label{TSP_eq2}
\text{IF}[n]=& \arg  \mathop{\min}_{p[n]\in \mathbf{P}} \bigg\{   \sum_{n=0}^{N-1} \mathcal{P}_1\Big\{\text{STFT}_s\big[n,p[n]\big]\Big\} \\
&+ \!\sum_{n=0}^{N-2}\! \mathcal{P}_2\Big\{p[n],p[n+1]\Big\}   \! + \! \sum_{s=1}^{S-1}\! \mathcal{P}_3\Big\{ g[s],g[s+1]\Big\} \!\bigg\},\notag
\end{align}
where $n\in[0,N-1]$ denotes the time index, and $N$ is the number of time-slices. $p[n]$ denotes all the possible paths belong to a set $\mathbf{P}$. Assuming every path can be approximated with a series of linear segments (red segments in Fig. \ref{fig3}), and $S$ is the number of IF segments. $g[s]$ denotes the gradient of the $s$-th segment. $\{\mathcal{P}_i\}_{ i=1,2,3}$ correspond to the three penalty functions, which are mathematically defined as follows:
\begin{align} 
\begin{cases}
 \mathcal{P}_1\Big\{\text{STFT}_s[n,f_j]\Big\} = j-1;  \notag\\
 \mathcal{P}_2\Big\{x,y\Big\} =
\begin{cases}
0, & \mid x-y\mid \leq \Delta_1; \\
c_1(\mid x-y\mid -\Delta_1), & \mid x-y\mid > \Delta_1;
\end{cases}\notag\\
 \mathcal{P}_3\Big\{x,y\Big\} = \begin{cases}
0, & \mid x-y\mid \leq \Delta_2; \\
c_2(\mid x-y\mid -\Delta_2), & \mid x-y\mid > \Delta_2.
\end{cases}\end{cases}
\end{align}

\begin{itemize}
\item $\mathcal{P}_1$:  $j\in [1,N_{\text{fft}}]$, and $N_{\text{fft}}$ is the number of  frequency samples at a certain time-slice. The values of $\text{STFT}_s[n,f]$ are sorted by a nonincreasing sequence: $\big|\text{STFT}_s[n,f_1]\big|\geq \big|\text{STFT}_s[n,f_2]\big| \cdots \geq \big|\text{STFT}_s[n,f_j]\big| \cdots \geq \big|\text{STFT}_s[n,f_{N_{\text{fft}}}]\big|$. This implies that the stronger the value of $\big|\text{STFT}_s[n,f_j]\big|$ is, a smaller value is assigned to the penalty function  $\mathcal{P}_1$.

\item $\mathcal{P}_2$:  In  traditional TF maxima detection methods, each IF point at a specific time-slice is independently detected from other time-slices, which omits the dependency among consecutive IF points. Since the IF law of an FM mode should be a smooth curve, we merge the maxima detection with the smooth constraint describing the transition rule of adjacent instants, which is mathematically quantified by the penalty parameter $c_1$ and the tolerable frequency change $\Delta_1$.

\item $\mathcal{P}_3$: This penalty function is mainly used to avoid estimation errors caused by IF intersections. As shown in Fig. \ref{fig3}, we calculate the curvature of every two adjacent IF segments, and ensure  correct IF estimation of signal modes by penalizing the combination of linear segments with large curvature. The penalty parameter $c_2$ and the tolerable curvature change $\Delta_2$ are determined according to the specific application at hand.
\end{itemize}

The procedure of the initial IF estimation method for multi-mode signals is summarized in \textit{Algorithm 1}. The determination of various parameters and the reasons for their choice will be discussed briefly in Section V. 

\IncMargin{.5em}
\begin{algorithm}[!t]
\caption{Initial IF Estimation.} \label{alg1}
\setstretch{1.}
\KwIn{$\text{STFT}_s$; $c_1,c_2,\Delta_1,\Delta_2,\delta, \varepsilon_0$;}

\textbf{Step 1}: Applying the optimal expression in (\ref{TSP_eq2}) to find the first optimal path, i.e., $\text{IF}^{(i)}[n],~i=0$, which corresponds to the strongest signal  mode; \\

\textbf{Step 2}: Generating a new STFT by setting zero value in the TF region around the detected optimal path by \textit{Step 1}, i.e., $\big[\text{IF}^{(i)}[n]-\delta,\text{IF}^{(i)}[n]+\delta\big]$ $\leftarrow 0$; \\
 
\textbf{Step 3}: If the energy in the remaining STFT is less than a predefined threshold value $\varepsilon_0$, terminate the procedure. Otherwise, repeat  \textit{Step 1} and \textit{Step 2} to obtain the next optimal path $\text{IF}^{(i)}[n]$,~ $i=i+1$;

\KwOut{   $\hat{f}_i[n]$ $\leftarrow \text{IF}^{(i)}[n],~~i=1,\cdots, I$.}
\end{algorithm}
\DecMargin{.5em}

\subsection{Enhanced IF Estimation}

Despite the effectiveness of \textit{Algorithm 1} for complicated signals, the resultant IF curves are probably unsmooth or fluctuating especially in strong noise environments. Inaccurate IF estimation is the main factor causing mode reconstruction error, i.e., random IF deviation from true IF trajectories would have a serious effect for recovering underlying signal modes, which motivates us to further enhance the initial IF information obtained in Section III.A.

In this subsection, the theoretical foundation, under which an iterative IF enhancement method is designed, is described. To begin with we provide a brief background on the sinusoidal time-frequency distribution \cite{VALEAU20041147,8894138Hua}, based on which a kernel phase averaging is explored for signal filtering and separation with low computational complexity. Detailed analysis on optimal filtering window and  separation interference  will be discussed in later section.

\subsubsection{Sinusoidal Time-Frequency Distribution (STFD)}

The STFD is particularly designed for a sinusoidal frequency modulated (SFM) signal $z[n]$ with a sinusoidal IF  $f[n]= \rho\cos(2\pi \mathsf{f} nT+\theta)$, defined as below:
\begin{align} \label{TSP_eq3}
z[n]&=\exp\bigg\{j2\pi T \sum_{m=0}^n \rho\cos(2\pi
\mathsf{f} mT+\theta)\bigg\}\notag\\&=\exp\bigg\{j\frac{\rho}{\mathsf{f}}\sin(2\pi \mathsf{f}
nT+\theta)\bigg\},
\end{align}
where $\mathsf{f}$ means a constant modulation frequency. The computation of STFD is based on a kernel function adapted to the SFM signal $z[n]$ \cite{VALEAU20041147}:
\begin{align} \label{TSP_eq4}
\text{STFD}_{z}[n,v]=\mathcal{F} \Big\{\mathcal{K}_{z}[n,l]\Big\},
\end{align}
where $n$ and $v$ denote time index and frequency index, $\mathcal{F}$ denotes the Fourier transform, and $\mathcal{K}_{z}[n,l]$ is the kernel function:
\begin{equation} \label{TSP_eq5}
\mathcal{K}_{z}[n,l]=\mathop{\Pi}_{p=1}^P
\Big\{z[n+l+a_p]^{b_p}z^*[n-l-a_p]^{b_p}\Big\},
\end{equation}
where $*$ denotes the conjugate operator. It is expected that the STFD can optimally concentrate the signal energy along its sinusoidal IF, i.e.,
\begin{equation} \label{TSP_eq6}
\mathcal{F} \Big\{\mathcal{K}_{z}[n,l]\Big\}=\delta\Big(v-\rho\cos(2\pi \mathsf{f} nT+\theta)\Big),
\end{equation}
which is equivalent to:
\begin{equation} \label{TSP_eq7}
\Theta \Big\{\mathcal{K}_{z}[n,l]\Big\}=2\pi lT  \cos(2\pi \mathsf{f} nT+\theta),
\end{equation}
where $\delta(\cdot)$ denotes the Dirac delta function, and $\Theta\{\cdot\}$ denotes the phase operator.
One of the solutions to (\ref{TSP_eq7}) can be obtained by using simple trigonometric relations when
 $P=2$, which gives
\begin{align} \label{TSP_eq8}
\begin{cases}
 a_1=0, \quad a_2=\nint[\big]{\frac{f_s}{4\mathsf{f}}};\\
 b_1=\pi \mathsf{f}lT \sin(2\pi \mathsf{f}lT), ~ b_2=\pi \mathsf{f}lT \cos(2\pi
\mathsf{f}lT);
\end{cases}
\end{align}
where $\nint{\cdot}$ denotes the rounding operator, and $f_s=\frac{1}{T}$ is the sampling rate. Therefore, one of the possible kernel functions for STFD in (\ref{TSP_eq4}) and (\ref{TSP_eq5}) is derived as below:
\begin{align} \label{TSP_eq9}
\mathcal{K}_{z}[n,l]&=\bigg\{z[n+l]z^*[n-l]\bigg\}^{\pi\mathsf{f}lT \sin(2\pi \mathsf{f}lT)}  \\
&\cdot \bigg\{z\Big[n+l+\nint[\big]{\frac{f_s}{4\mathsf{f}}}\Big]z^*\Big[n-l-\nint[\big]{\frac{f_s}{4\mathsf{f}}}\Big]\bigg\}^{\pi\mathsf{f}lT
\cos(2\pi \mathsf{f}lT)}.\notag
\end{align}

It is noted that the kernel $\mathcal{K}_{z}[n,l]$ of  STFD is closely dependent on the modulation frequency $\mathsf{f}$, and the exponent of $\mathcal{K}_{z}[n,l]$ is a function of $\mathsf{f}$ and the time lag $l$, indicating that the parameter $\mathsf{f}$ should be known in advance or estimated to compute  STFD. Details of the above derivations could be found in \cite{VALEAU20041147} and are not presented here for brevity. This technique has also been applied to enhance electric network signal, a mono-mode signal, for forensic analysis \cite{8894138Hua}. Here, we consider a more complicated multi-mode scenario.

\subsubsection{Signal Encoding}

The purpose of introducing  STFD is to explore its property for enhancing initial IF estimation of the input $s[n]$ in Fig. \ref{fig2}. By comparing the signal $s[n]$  in (\ref{TSP_eq1}) and the signal $z[n]$ adapted to STFD in (\ref{TSP_eq3}), we attempt to transform the sinusoidal modulated waveform $s[n]$  into an approximate SFM signal by encoding $s[n]$  into the IF of a unit-amplitude analytic signal:
\begin{align} \label{TSP_eq10}
z_s[n]&=\exp\bigg\{j2\pi T\mu\sum_{m=0}^n s[m]\bigg\}\notag\\
&=\exp\bigg\{j2\pi T\mu\sum_{m=0}^n \sum_{i=1}^{I}x_i[m] \bigg\} \exp\bigg\{j2\pi T\mu\sum_{m=0}^n \epsilon[m]\bigg\}\notag\\&=z_x[n] z_{\epsilon}[n],
\end{align}
where $\mu$ denotes the modulation index, $z_{\epsilon}[n]=\exp\big\{j2\pi T\mu\sum_{m=0}^n \epsilon[m]\big\}$ denotes the encoded noise, which becomes multiplicative relationship with the encoded clean signal $z_x[n]=\exp\big\{j2\pi T\mu\sum_{m=0}^n \sum_{i=1}^{I}x_i[m] \big\}$.  

By applying the STFD in (\ref{TSP_eq4}) on the encoded signal $z_s[n]$, the underlying signal ${x}[n]=\sum_{i=1}^{I}{x}_i[n]$ is obtained by estimating the IF of the STFD on $z_s[n]$, which could be realized by detecting the peaks of  $\text{STFD}_{z_s}[n,v]$ at all the time-slices. However, it is cumbersome for this purpose since in practical applications: \textbf{i)}   the implementation of $\text{STFD}_{z_s}[n,v]$ is expensive due to the discrete Fourier transform (DFT) on the nonlinear kernel function; \textbf{ii)} even though the underlying signal ${x}[n]$ is recovered, it is still hard to decompose it into individual modes. Therefore, a low complexity method allowing for mode separation is designed, as discussed in the next part.

\subsubsection{Kernel Phase Averaging}

This method, referred to as kernel phase averaging (KPA), has two attractive features. Instead of finding the STFD followed by peak search, the IF estimation of $z_s[n]$ can be realized by a simple phase operation on kernel function without implementing the complex STFD; More importantly, the underlying signal modes $\{{x}_i[n]\}_{i=1,\cdots,I}$ can be individually separated by providing initial IFs of the signal $s[n]$, which gives rise to enhanced IF of each mode. We then present the following theorems of the KPA method as well as their proofs. \vspace{1.8mm}
\\\textbf{{Theorem 1}} ~ \textit{For a noisy mono-mode signal: $s[n]=x[n]+\epsilon[n]=\rho\cos \Big(2\pi  \mathsf{f}[n] nT+\theta \Big)+\epsilon[n],~n=0,\cdots,N-1$, the kernel phase averaging on STFD's kernel of $z_s[n]$ theoretically equals to the underlying clean signal $x[n]$, provided that the IF of $s[n]$ approximates to a constant within a certain range:
\begin{align} \label{TSP_eq11}
\mathbb{E}\bigg\{\frac{\Theta\big\{\mathcal{K}_{z_s}[n,l]\big\}\big|_{\mathsf{f}[n]}}{2\pi l T\mu}\bigg\}
\!\approx \! \rho\cos \Big(2\pi  \mathsf{f}[n] n T+\theta \Big) \!=\! x[n],
\end{align}
where $\mathbb{E}\{\cdot\}$ denotes the expectation operator, and $ \mathcal{K}_{z_s}$ denotes the kernel function of STFD on $z_s[n]$. $\mathsf{f}[n]$ is the IF at the $n$-th time-slice, which is regarded as a constant within $[n-L/2,n+L/2]$, where $L$ is the window length of the KPA method.
}

\textit{Proof of Theorem 1.} The exponent of the kernel function in (\ref{TSP_eq9}) is dependent on the time lag $l$, which allows for a direct estimation of the underlying signal $x[n]$ by performing an averaging operation on the phase of the kernel in (\ref{TSP_eq9}) by multiplying the term $\frac{1}{2\pi l T\mu}$:
\begin{align} \label{TSP_eq12}
&\frac{\Theta\big\{\mathcal{K}_{z_s}[n,l]\big|_{\mathsf{f}[n]}\big\}}{2\pi l T\mu}
\\&= \Theta\bigg\{
\mathop{\Pi}_{i=1}^2\big\{z_s[n+l+a_{i}]\big\}^{\frac{b_{i}}{2\pi l T\mu}}\big\{z_s^*[n-l-a_{i}]\big\}^{\frac{b_{i}}{2\pi l T\mu}}\bigg\} \nonumber, 
 \end{align}
where $l\in [-L/2,L/2]$.  By encoding $s[n]$ into $z_s[n]$ according to (\ref{TSP_eq10}), and then bringing $z_s[n]$ into (\ref{TSP_eq12}), the expectation operation permits an unbiased signal recovery of $x[n]$ from the noisy signal $s[n]$, derived as below:
\begin{align}  \label{TSP_eq13}
&\mathbb{E}\bigg\{\frac{\Theta\big\{\mathcal{K}_{z_s}[n,l]\big|_{\mathsf{f}[n]}\big\}}{2\pi l T\mu}\bigg\}
=\mathbb{E}\bigg\{\Theta \Big\{ \mathcal{K}_{z_x}[n,l]\cdot \mathcal{K}_{z_{\epsilon}}[n,l]\Big\}\bigg\}\notag \\
=&\mathbb{E}\bigg\{ \Theta\Big\{ \mathcal{K}_{z_x}[n,l]\Big\}+\Theta\Big\{ \mathcal{K}_{z_{\epsilon}}[n,l]\Big\}\bigg\},
\end{align}
where
\begin{align} 
\begin{cases}
\mathcal{K}_{z_{x}}[n,l]
= \exp\bigg\{j\pi T \mathsf{f}[n]  \Big\{\sin(2\pi \mathsf{f}[n] lT)\sum_{n-l}^{n+l}x[\lambda]\\
\qquad\qquad\qquad\qquad +\cos(2\pi \mathsf{f}[n] lT)\sum_{n-l-\nint{\frac{f_s}{4\mathsf{f}[n]}}}^{n+l+\nint{\frac{f_s}{4\mathsf{f}[n]}}}x[\lambda]\Big\}\bigg\}; \notag\\
\mathcal{K}_{z_{\epsilon}}[n,l]
=\exp\bigg\{j\pi T \mathsf{f}[n]  \Big\{\sin(2\pi \mathsf{f}[n] lT)\sum_{n-l}^{n+l}\epsilon[\lambda] \\
\qquad\qquad\qquad\qquad +\cos(2\pi \mathsf{f}[n] lT)\sum_{n-l-\nint{\frac{f_s}{4\mathsf{f}[n]}}}^{n+l+\nint{\frac{f_s}{4\mathsf{f}[n]}}} \epsilon[\lambda]\Big\}\bigg\}.\notag
\end{cases}
\end{align}
Under the assumption that $\epsilon(t)$ is zero-mean additive noise and $x[n]=\rho\cos\big(2\pi  \mathsf{f}[n] nT+\theta\big)$,  we have $\mathbb{E}\Big\{ \Theta\big\{ \mathcal{K}_{z_{\epsilon}}[n,l]\big\}\Big\}\approx 0$, therefore
\begin{align} \label{TSP_eq14}
&\mathbb{E}\bigg\{\frac{\Theta\big\{\mathcal{K}_{z_s}[n,l]\big\}}{2\pi l T\mu}\bigg\}
\approx \mathbb{E}\bigg\{ \Theta\Big\{ \mathcal{K}_{z_x}[n,l]\Big\}\bigg\}\notag\\
=&A  \bigg\{\!\sin(2\pi \mathsf{f}[n] lT)\sum_{n-l}^{n+l}x[\lambda] \!+\!\cos(2\pi \mathsf{f}[n] lT)\!\sum_{n-l-\nint{\frac{f_s}{4\mathsf{f}[n]}}}^{n+l+\nint{\frac{f_s}{4\mathsf{f}[n]}}} x[\lambda]\!\bigg\} \notag\\
=&\rho\cos(2\pi \mathsf{f}[n] nT +\theta)
= x[n],
\end{align} 
where $A=\pi T  \mathsf{f}[n]$. Note that the phase term $\Theta\big\{ \mathcal{K}_{z_x}[n,l]\big\}$ on $z_x[n]$ keeps constant over the time lag $l$, which gives a precise estimation of the clean signal $x[n]$. In the presence of noise, $\Theta\big\{\mathcal{K}_{z_s}[n,l]\big\}$ on $z_s[n]$ changes with the variation of $l$, under this circumstance, the underlying signal $x[n]$ can be recovered by taking the mean operation on the fluctuated phase term $\Theta\big\{\mathcal{K}_{z_s}[n,l]\big\}$ over the time lag $l$.  The proof of Theorem 1 is completed. \QEDA
\vspace{1.8mm}
\\\textbf{{Theorem 2}} ~ \textit{For a noisy multi-mode signal with $I$ modes: $s[n]=\sum_{i=1}^{I} x_i[n]+\epsilon[n]=\sum_{i=1}^{I} \rho_i\cos \Big(2\pi  \mathsf{f}_i[n] nT+\theta_i \Big)+\epsilon[n],~n=0,\cdots,N-1$, the kernel phase averaging on STFD's kernel of $z_s[n]$ theoretically equals to the underlying clean mode $x_i[n]$, provided that \textbf{i)} the IF of the $i$-th mode inserted into the kernel function approximates to a constant within a certain range; \textbf{ii)}  a separation condition in (\ref{TSP_eq27}) should be satisfied. Under the above conditions, we have
\begin{align} \label{TSP_eq15}
\hat{x}_i[n]=&\mathbb{E}\bigg\{\frac{\Theta\big\{\mathcal{K}_{z_s}[n,l]\big\}\big|_{\mathsf{f}_i[n]}}{2\pi lT\mu}\bigg\}\notag\\
\approx&  \rho_i\cos \Big(2\pi  \mathsf{f}_i[n] nT+\theta_i \Big)=  x_i[n], 
\end{align}
where $\mathsf{f}_i[n]$ is the IF of the $i$-th mode at the $n$-th time-slice, which is regarded as a constant within $[n-L/2,n+L/2]$. Consequently, the clean multi-mode signal $x[n]$ can be recovered through:
\begin{align} \label{TSP_eq16}
\hat{x}[n]&= \sum_{i=1}^{I}\mathbb{E}\bigg\{\frac{\Theta\big\{\mathcal{K}_{z_s}[n,l]\big\}\big|_{\mathsf{f}_i[n]}}{2\pi lT\mu}\bigg\}\notag\\
&\approx \sum_{i=1}^{I} \rho_i\cos \Big(\!2\pi  \mathsf{f}_i[n] nT+\theta_i \!\Big)=\sum_{i=1}^{I}  x_i[n].
\end{align}
}

\textit{Remark.}  According to (\ref{TSP_eq15}) and (\ref{TSP_eq16}), it is worth stressing that the $i$-th signal mode, $x_i[n]$,  can be successfully reconstructed if its IF, $\mathsf{f}_i[n]$, is known or estimated. Intuitively, the reconstructed $\hat{x}_i[n]$ is inevitably influenced by other signal modes, i.e., $\sum_{j \neq i}^{I}x_j[n]$, especially when the IF laws $\{\mathsf{f}_i[n]\}_{i=1,\cdots,I}$ are closely-spaced or crossing. The proof of \textit{Theorem 2} will be provided in Section IV. A.

\subsubsection{Iterative IF Refinement}

Each signal mode can be filtered by the operation in (\ref{TSP_eq15}), and meanwhile the noise variance is greatly mitigated. Thus, the initial IF estimation in Section III.A is further enhanced based on STFT of estimated $\{\hat{x}_i[n]\}_{i=1,\cdots,I}$ by detecting STFT peaks across each time index \cite{Boashash135378}:
\begin{align} \label{TSP_eq17}
\tilde{f}_i[n]\! = \! \frac{ \arg  \mathop{\max}_{v} \Big\{\text{STFT}_{\hat{x}^{\prime}_i[n]}[n,v]\Big\} }{TN_{\text{fft}}}, ~~ i=1,\cdots,I
\end{align}
where $N_{\text{fft}}$ is the DFT length of STFT computation, and $\hat{x}^{\prime}_i[n]$ denotes the analyzed version of $\hat{x}_i[n]$.

It has been demonstrated that one iteration would be sufficient for a desirable enhancement performance. However, the enhanced signal modes using initial IFs can offer an updated IF estimate, which means that the KPA method can be implemented in an iterative manner, i.e.,  (\ref{TSP_eq15}) in  \textit{Theorem 2} is repeatedly conducted by updating  IF information until a satisfied performance is achieved or a prescribed number of iterations is reached.

\subsection{Combining IF and IA  for Mode Decomposition}

After obtaining the IF information of each mode, the subsequent task is the IA recovery, so that the mode reconstruction can be realized by combining  extracted IFs and IAs. Despite the effectiveness on IF enhancement using the KPA method, the IA of enhanced $\hat{x}[n]$ in (\ref{TSP_eq16}) might be severely distorted. It would become more difficult to recover the IA as the SNR level decreases. In our study, the distortion of mode's IA  can be compensated by taking into account the information in linear STFT, which is a commonly-used linear TFR with a complex-valued TF description that involves magnitude contribution, thus enabling IA  reconstruction.

On the one hand, the Gaussian window based STFT is explored for reconstructing the IA of each mode:
\begin{align} \label{TSP_eq18}
\text{STFT}_s[n,v]=\sum_{l=0}^{N-1}s^{\prime}[l]g[l-n]\exp\Big\{\frac{-j2 \pi  l v}{N}\Big\},
\end{align}
where $s^{\prime}[n]$ denotes the analyzed version of $s[n]$, and $g[n]$ is the real Gaussian window: $g[n]=\sigma^{-1} \exp\{\frac{-\pi n^2T^2}{\sigma^2}\}$, where $\sigma$ controls the Gaussian window size. 

On the other hand, according to the enhanced IFs $\{\tilde{f}_i[n]\}_{i=1,\cdots,I}$ in (\ref{TSP_eq17}), the synchroextracting operator (SEO) of our proposed ETFR-MD is defined as below:
\begin{align} \label{TSP_eq19}
\text{SEO}_{f}[n,v]=\sum_{i=1}^I \delta\Big(v-\nint[\Big]{\tilde{f}_i[n]TN_{\text{fft}}}\Big).
\end{align}

Based on the idea of the SET \cite{yu2017synchroextracting}, the  proposed ETFR associated with STFT is formulated using the SEO in (\ref{TSP_eq19}), expressed as:
\begin{align} \label{TSP_eq20}
\text{ETFR}_s[n,v]=\text{STFT}_s[n,v]\odot \text{SEO}_{f}[n,v],
\end{align}
where $\odot$ denotes the element-wise product operator. Note that the Gaussian window based  STFT achieves the maximum values along IF laws, thus the extracted IAs through (\ref{TSP_eq20}) will have the best performance of noise robustness.

After  obtaining the enhanced $\text{ETFR}_s[n,v]$, the $i$-th signal mode can be reconstructed by integrating $\text{STFT}_s[n,v]$ in the IF trajectory of the corresponding mode, i.e., using only the STFT coefficients on the enhanced IF ridge of each mode:
\begin{align} \label{TSP_eq21}
\tilde{x}_i[n]&=\Re\bigg\{\text{ETFR}_s\bigg[n,\nint[\Big]{\tilde{f}_i[n]TN_{\text{fft}}}\bigg]\bigg\}\notag\\
&=\Re\bigg\{ \text{STFT}_s[n,v]\cdot \delta\Big(v-\nint[\Big]{\tilde{f}_i[n]TN_{\text{fft}}}\Big)\bigg\},
\end{align}
where $i=1,\cdots, I$ and $\Re\{\cdot\}$ means the real part of a complex number. 
It also enables the signal enhancement in a straightforward manner, i.e., $\tilde{x}[n]=\sum_{i=1}^I \tilde{x}_i[n]$, which can be alternatively derived by:
\begin{align} \label{TSP_eq22}
\tilde{x}[n]=\Re\bigg\{\sum_{i=1}^I\text{ETFR}_s\bigg[n,\nint[\Big]{\tilde{f}_i[n]TN_{\text{fft}}}\bigg]\bigg\}.
\end{align}

\textit{Remark.} Signal energy of an ideal TFR will be only concentrated on true IF trajectories. Nevertheless, the utilization of STFT will inevitably give rise to biased IF estimation due to the non-linearity of some modes, which further leads to mode reconstruction error. Regarding the use of STFT for both initial IF estimation and IA extraction, one of the key issues is to determine an appropriate choice of the Gaussian window size  in (\ref{TSP_eq18}). In order to minimize the impact on IF estimation and mode decomposition, we will choose empirical values via experiments like in \cite{Pham7885114,Yu8458385,yu2017synchroextracting}. 

Returning to the problem posed at Section II, we now state that the goal to recover  IF and IA information of each mode is achieved, and the proposed ETFR-MD method is summarized in \textit{Algorithm 2}, where $K$ denotes the number of iterations for the KPA method. We next provide the theoretical analysis of the proposed method in Section IV, and validate the theoretical analysis by experimental results in  Section V. 

\IncMargin{.5em}
\begin{algorithm}[!t]
	\caption{Enhanced TFR and MD.} \label{alg2}
	\setstretch{1.}
	\KwIn{$s[n]$, $K$;}
	Compute $\text{STFT}_s[n,v]$ via (\ref{TSP_eq18})\;
	Initial IF estimation: $\{\hat{f}_i[n]\}_{i=1,\cdots,I} \leftarrow$ \textbf{Algorithm 1}\;
	Encode $s[n]$ via (\ref{TSP_eq10}), and we obtain $z_s[n]$\;
    	\For{$k=1, k \le K$}
    	{
    	Generate $\mathcal{K}_{z_s}[n,l]$ using updated IFs via (\ref{TSP_eq12})\;
    	Enhance $s[n]$ using KPA via (\ref{TSP_eq15}) and (\ref{TSP_eq16})\;
    Obtain updated IFs $\{\tilde{f}_i[n]\}_{i=1,\cdots,I}$ via (\ref{TSP_eq17})\;
    	}   
    	Compute $\text{SEO}_{f}[n,v]$ via  (\ref{TSP_eq19})\;
    	Enhanced TFR via  (\ref{TSP_eq20}) to obtain $\text{ETFR}_s[n,v]$ \;
    	Mode decomposition via  (\ref{TSP_eq21}) to obtain $\tilde{x}_i[n]$\;
\KwOut{   $\text{ETFR}_s[n,v]$ and ${\tilde{x}_i}[n], ~ i=1,\cdots,I$.}
\end{algorithm}
\DecMargin{.5em}

\section{Performance Analysis}

In this section, we first analyze that under what conditions \textit{Theorem 2} for mode enhancement holds. In this regard, a closed-form expression as a function of frequency difference is derived, which indicates the negligible mutual interference between different modes, as long as the frequency difference is below a certain limit. In addition, the optimal window length $L$ of the KPA method in (\ref{TSP_eq12}) used for IF enhancement is discussed, which is useful in analyzing how to adjust an appropriate window length according to the property of different modes.

\subsection{Interference Analysis of Mode Separation}

\begin{figure*}[!t]
\hspace{-6mm}
\subfigure[AAEs vs. $f_{\Delta}$ and Window Length $L$.]{
\includegraphics[width=3.8 in]{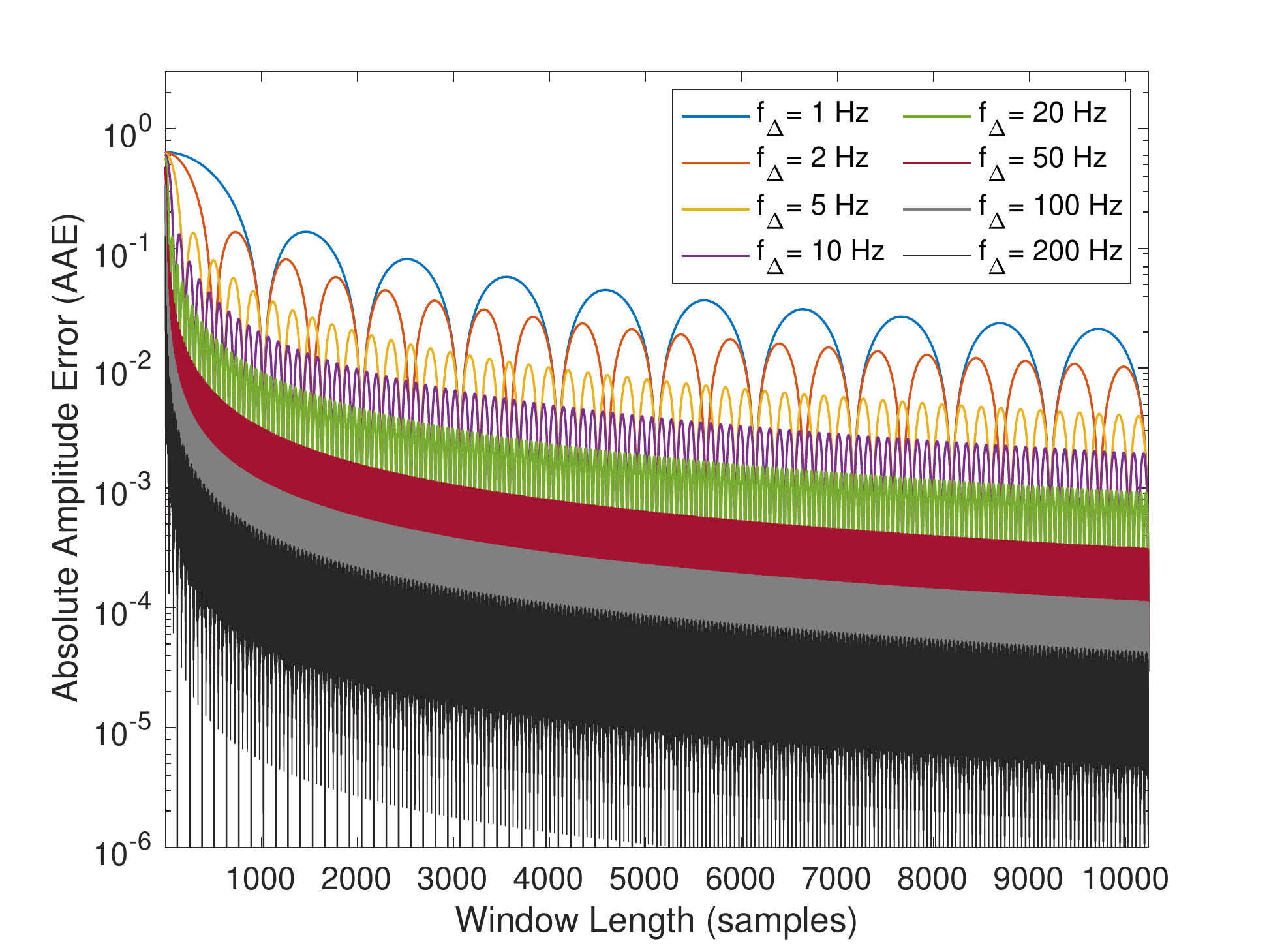}
}
\hspace{-7mm}
\subfigure[Averaged AAEs  vs. $f_{\Delta}$ and Window Length $L$.]{
\includegraphics[width=3.8 in]{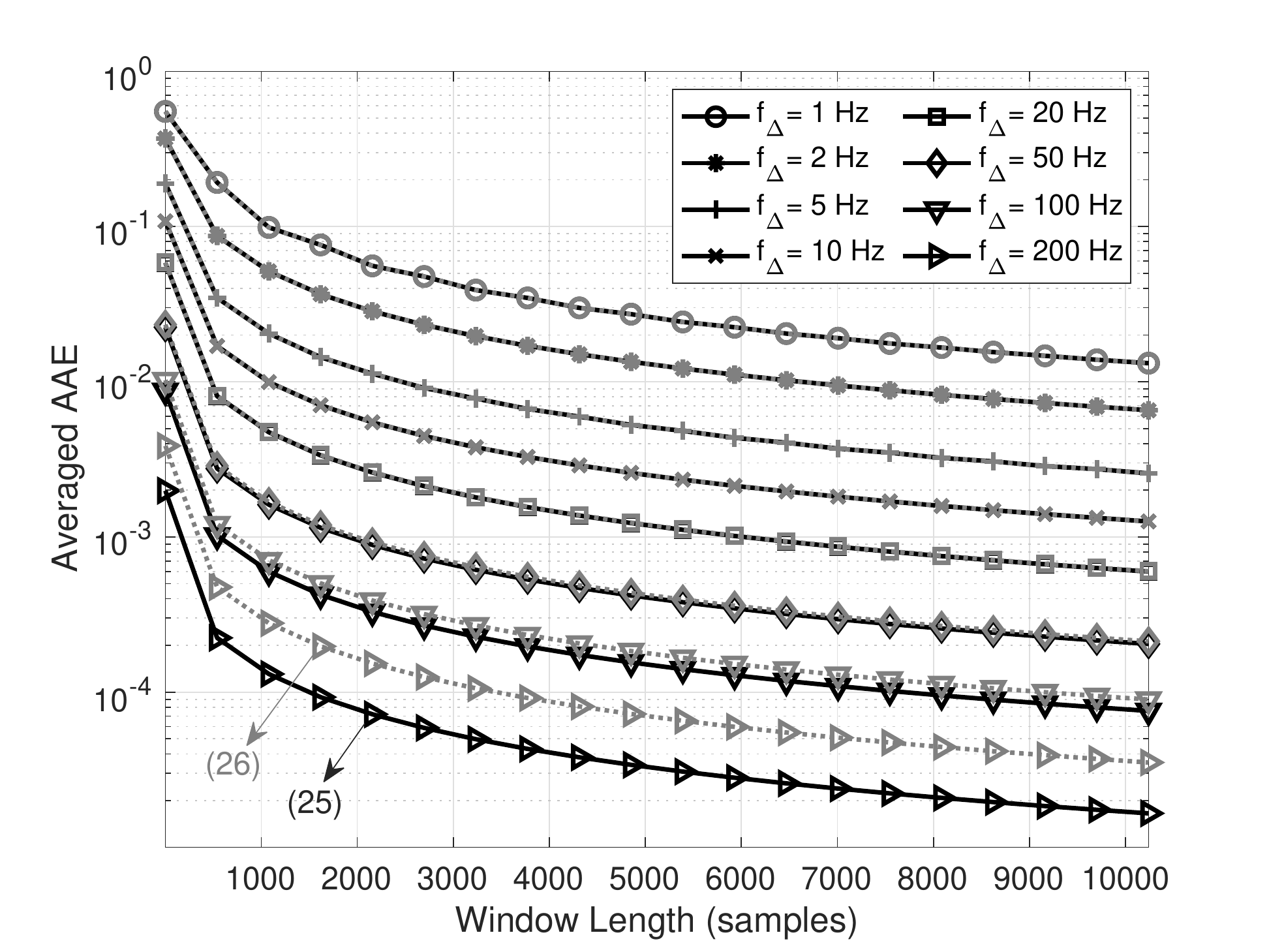}
}
\caption{Performance analysis of mode separation using the KPA method with $I=2$, $\rho_1=\rho_2=1$, $\mathsf{f}_1[n]\equiv$ 200 Hz, $\mathsf{f}_2[n]=\mathsf{f}_1[n]+f_{\Delta}$, $f_s$ = 1024 Hz, and $N=10240$. \textbf{(a)} Absolute amplitude errors (AAEs) of reconstructed signal mode computed from (25) versus different frequency difference $f_{\Delta}$ and window length $L$. \textbf{(b)} Averaged AAEs of reconstructed signal mode on the results in (a) with $N_a=256$ versus different frequency difference $f_{\Delta}$ and window length $L$. The black curves and gray curves are computed from (\ref{TSP_eq25}) and  (\ref{TSP_eq25b}), respectively.}
\label{fig4}
\end{figure*}

The closed-form expression of  mutual interference for mode separation via the KPA Method is firstly derived, and then an illustrative example is provided to verify the theoretical expression.

Recall \textit{Theorem 2} in Section III.B, when the received signal of interest has multiple modes, the KPA method can be used for mode separation according to (\ref{TSP_eq15}). In the following, we quantitatively analyze the cross-interference between signal modes by using KPA, and derive a condition under which the mode separation in (\ref{TSP_eq15}) can be well carried out. In order to avoid bias, we assume a multi-mode signal $s[n]$, each mode of which has approximately constant IF within the window range of the KPA method. According to (\ref{TSP_eq10}), we obtain the encoded multi-mode signal $z_s[n]=\exp\Big\{j2\pi\mu T \sum_{m=0}^n \sum_{i=1}^I\rho_{i}\cos\big(2\pi \mathsf{f}_i[n] mT+\theta_{i}\big)\Big\}$. 
To prove \textit{Theorem 2}, it is equivalent to proving the validity of the following expression:
\begin{align}  \label{TSP_eq23}
\mathbb{E}\bigg\{\frac{\Theta\big\{\mathcal{K}_{z_s}[n,l]\big\}\big|_{\mathsf{f}_i[n]}}{2\pi lT\mu}\bigg\} \approx   x_i[n], ~~ i=1,\cdots, I.
\end{align}

\textit{Proof of Theorem 2.}  In a straightforward manner, the $i$-th mode in (\ref{TSP_eq16}) is calculated as below:
\begin{align} \label{TSP_eq24}
& \mathbb{E}\bigg\{\frac{\Theta\big\{\mathcal{K}_{z_s}[n,l]\big\}\big|_{\mathsf{f}_i[n]}}{2\pi lT\mu}\bigg\}=\mathbb{E}\bigg\{ \Theta\Big\{ \mathcal{K}_{z_x}[n,l]\Big\}\bigg\} \notag \\
= & \mathbb{E}\bigg\{\frac{ b_{1i}}{\mathsf{f}_i[n]} \rho_{i}\cos\big(2\pi \mathsf{f}_i[n] nT+\theta_{i}\big)\sin\big(2\pi \mathsf{f}_i[n]T(l+a_{1i})\big)\notag\\&
 +\frac{b_{2i}}{\mathsf{f}_i[n]} \rho_{i}\cos\big(2\pi \mathsf{f}_i[n]nT+\theta_{i}\big)\sin\big(2\pi \mathsf{f}_i[n]T(l+a_{2i})\big) \notag \\
&+\!\sum_{i'\neq i}^I \!\frac{ b_{1i}}{\mathsf{f}_{i'}[n]} \rho_{i'}\cos\big(2\pi \mathsf{f}_{i'}[n]nT\!+\!\theta_{i'}\big)\sin\big(2\pi \mathsf{f}_{i'}[n]T(l\!+\! a_{1i})\big)\notag\\&
+\frac{b_{2i}}{\mathsf{f}_{i'}[n]} \rho_{i'}\cos\big(2\pi \mathsf{f}_{i'}[n]nT+\theta_{i'}\big)\sin\big(2\pi \mathsf{f}_{i'}[n]T(l+a_{2i})\big) \bigg\} \notag\\
=& \!\sum_{i'\neq i}^I\! \rho_{i'}\! \cos\big(2\pi \mathsf{f}_{i'}[n]nT+\theta_{i'}\big) \mathbb{E}\!\bigg\{ \! \frac{b_{1i}}{\mathsf{f}_{i'}[n]} \! \sin\! \big(2\pi \mathsf{f}_{i'}[n]T(l\!+\! a_{1i})\big)\notag\\&
 +\frac{b_{2i}}{\mathsf{f}_{i'}[n]}  \sin\big(2\pi \mathsf{f}_{i'}[n]T(l+a_{2i})\big) \bigg\}+x_i[n]\notag\\
 = & \text{INT}\big(\mathsf{f}_{i}[n],\mathsf{f}_{i'}[n]\big)+x_i[n],
\end{align}
where $a_{1i}=0$, $a_{2i}=\nint[\big]{\frac{f_s}{4\mathsf{f}_{i}[n]}}$, $b_{1i}= \mathsf{f}_{i}[n]  \sin(2\pi
\mathsf{f}_{i}[n] lT)$, and $ b_{2i}= \mathsf{f}_{i}[n]  \cos(2\pi \mathsf{f}_{i}[n] lT)$, which are derived according to (\ref{TSP_eq8}). $\text{INT}\big(\mathsf{f}_{i}[n],\mathsf{f}_{i'}[n]\big)$ denotes the quantity of interference on the $i$-th mode from other modes:
\begin{align} \label{TSP_eq25}
&\text{INT}\big(\mathsf{f}_{i}[n],\mathsf{f}_{i'}[n]\big)\Big| _{l\in [-L/2,L/2]}\notag\\=&
\sum_{i'\neq i}^I  A_{i'} \mathbb{E} \bigg\{\frac{\mathsf{f}_{i}[n]}{\mathsf{f}_{i'}[n]}   \sin\big(2\pi
\mathsf{f}_{i}[n]lT)  \sin\!\big(2\pi \mathsf{f}_{i'}[n]lT\big)
 \\& +\frac{\mathsf{f}_{i}[n]}{\mathsf{f}_{i'}[n]}   \cos\big(2\pi\mathsf{f}_{i}[n] lT\big) \sin\Big(2\pi \mathsf{f}_{i'}[n]T\big(l+\nint[\big]{\frac{f_s}{4\mathsf{f}_{i}[n]}}\big)\Big) \bigg\},\notag
\end{align} 
where $A_{i'}=\rho_{i'}\cos\big(2\pi \mathsf{f}_{i'}[n]nT\!+\!\theta_{i'}\big)$, and we define the frequency difference $f_{\Delta}[n]=\big|\mathsf{f}_{i}[n]-\mathsf{f}_{i'}[n]\big|$. Note  that when $\mathsf{f}_{i}[n] \approx \mathsf{f}_{i'}[n]$, the expression in (\ref{TSP_eq25}) is simplified into:
\begin{align} \label{TSP_eq25b}
\text{INT}\big(\mathsf{f}_{i}[n],\mathsf{f}_{i'}[n]\big)
{\approx} \sum_{i'\neq i}^I  A_{i'} \frac{\mathsf{f}_{i}[n]}{\mathsf{f}_{i'}[n]} \mathbb{E}\bigg\{\!\cos\Big(\! 2\pi 
f_{\Delta}[n] lT \!\Big)\!\bigg\}. 
\end{align} 

It is seen from (\ref{TSP_eq25}) and (\ref{TSP_eq25b}) that the interference level is mainly dependent on the frequency difference $f_{\Delta}$ between the mode of interest and other ones, while also dependent on the window length $L$. An illustrative example is given in Fig. \ref{fig4}, where  Fig. \ref{fig4} (a) shows the  absolute amplitude errors (AAEs) computed by (\ref{TSP_eq25}) versus different values of $L$ and $f_{\Delta}$ for a two-mode signal separation with $\mathsf{f}_1[n]\equiv$  200 Hz, $f_s$ = 1024 Hz, and $N=10240$. As functions of window length $L$, all the AAE curves present a trend of periodic oscillation. Fig. \ref{fig4} (b) shows the averaged AAEs  by computing the mean values of every $N_a=256$ samples in Fig. \ref{fig4} (a), where the black curves indicate the mean AAEs computed by (\ref{TSP_eq25}), while the gray curves are the counterparts computed by (\ref{TSP_eq25b}). It verifies the approximate expression from (\ref{TSP_eq25}) to (\ref{TSP_eq25b}) when $\mathsf{f}_{1}[n] \approx \mathsf{f}_{2}[n]$, i.e., a small $f_{\Delta}[n]$. Although the expression (\ref{TSP_eq25b})  does not hold for a large $f_{\Delta}[n]$, the degree of interference is greatly reduced as the frequency difference  increases, e.g., $f_{\Delta}[n]$ = 200 Hz in Fig. \ref{fig4} (b). It is observed that the mean interference becomes insignificant with a large $L$ and $\Delta_f$, which means that the interference can be neglected when the spectral contents of multiple signal modes are sufficiently separated.

Based on the derived expression from (\ref{TSP_eq24}) to (\ref{TSP_eq25b}) for multi-mode signals, it is concluded that one mode can be well separated from other modes when the interference term $\text{INT}\big(\mathsf{f}_{i}[n],\mathsf{f}_{i'}[n]\big)$ in (\ref{TSP_eq25}) satisfies the following condition:
\begin{align} \label{TSP_eq27}
\bigg|\mathbb{E}\bigg\{\cos\Big(2\pi 
f_{\Delta}[n] lT \Big)\bigg\}\bigg| _{l\in [-L/2,L/2]} < \varepsilon_1,
\end{align} 
where $\varepsilon_1$ is a small threshold value which denotes the allowed interference level. Under this condition, the expression (\ref{TSP_eq16}) is verified by summing all the mode terms of (\ref{TSP_eq24}). \QEDA

\textit{Remark.} It should be stressed that the KPA method focuses on the IF enhancement of multiple signal modes, and we are less concerned with the effect of the frequency difference $f_{\Delta}[n]$ on signal amplitude.  It implies that a significant amplitude interference in (\ref{TSP_eq24})  will cause weaker frequency interference instead, i.e., the IF information of signal modes may still be restored correctly, although it has received a certain degree of amplitude distortion. Besides, Fig. \ref{fig4} demonstrates that the larger the window length $L$ is, the less interference is generated. However, this conclusion is based on the assumption that the IFs of modes remain constant in a limited range, i.e., $\{\mathsf{f}_i[n]\}_{i=1,\cdots,I}\equiv 200,~ n\in[0,\cdots,N-1]$.  Intuitively, the KPA method becomes biased when this assumption does not hold, meaning that the window length cannot be increased indefinitely. This naturally leads to the question of the optimal selection of $L$, which will be analyzed in next subsection.

\subsection{Optimal Window Length of IF Enhancement}

\setcounter{equation}{29}
\begin{figure*}[!h]
\begin{align} \label{TSP_eq30}
&\hat{x}[n]=\mathbb{E}\Bigg\{\frac{\Theta\big\{\mathcal{K}_{z_x}[n,l]\big\}\big|_{\hat{f}[n]}}{2\pi lT}\Bigg\}\\ &= 
\mathbb{E} \Bigg\{\sum_{p=1}^{2} b_p
\cdot \Big\{ \frac{\sin\big(2\pi T \mathsf{f}_0(n+l+a_p)+\pi T^2 \mathsf{r}_0(n+l+a_p)^2 \big)}{\mathsf{f}_0 +\mathsf{r}_0 T(n+l+a_p)}
-\frac{\sin\big(2\pi T \mathsf{f}_0(n-l-a_p)+\pi T^2 \mathsf{r}_0(n-l-a_p)^2 \big)}{\mathsf{f}_0 +\mathsf{r}_0 T(n-l-a_p)}  \Big\}\Bigg\}\notag\\
& =  \mathbb{E}\Bigg\{ \frac{\Big(\mathsf{f}_0+\mathsf{r}_0 nT\Big)\hat{f}[n]\sin\Big(2\pi lT \hat{f}[n]\Big)\sin\Big(2\pi lT(\mathsf{f}_0+\mathsf{r}_0 nT)\Big)\cos\Big(2\pi \mathsf{f}_0 nT +\pi \mathsf{r}_0 T^2 (n^2+l^2)\Big)}{\Big(\mathsf{f}_0+\mathsf{r}_0 nT+\mathsf{r}_0 lT\Big)\Big(\mathsf{f}_0+\mathsf{r}_0 nT -\mathsf{r}_0 lT\Big)} \notag\\ & 
\quad\quad  +\frac{\Big(\mathsf{f}_0+\mathsf{r}_0 nT\Big)\hat{f}[n]\cos\Big(2\pi lT \hat{f}[n]\Big)\sin\Big(2\pi T(l+\nint{\frac{f_s}{4\hat{f}[n]}})(\mathsf{f}_0+\mathsf{r}_0 nT)\Big)\cos\Big(2\pi \mathsf{f}_0 nT +\pi  \mathsf{r}_0 T^2  \big(n^2+(l+\nint{\frac{f_s}{4\hat{f}[n]}})^2\big)\Big)}{\Big(\mathsf{f}_0+\mathsf{r}_0 nT+\mathsf{r}_0 lT+\mathsf{r}_0 T \nint{\frac{f_s}{4\hat{f}[n]}}\Big)\Big(\mathsf{f}_0+\mathsf{r}_0 nT -\mathsf{r}_0 lT-\mathsf{r}_0 T\nint{\frac{f_s}{4\hat{f}[n]}}\Big)}\Bigg\}.\notag
\end{align}
\rule{\linewidth}{.5pt}
\end{figure*}
\setcounter{equation}{27}

As illustrated in Fig. \ref{fig4} and (\ref{TSP_eq27}), due to the periodic oscillation of trigonometric function, the inter-mode interference drops sharply to zero as long as the window length $L$ satisfies the following condition:
\begin{align} \label{TSP_eq28}
\Big\{f_{\Delta}[n] L T \Big\} \in \mathbb{Z}~ \Longrightarrow~ 
L = \frac{n_0 }{f_{\Delta}[n] T} = n_0 \frac{ f_s}{f_{\Delta}[n]},
\end{align} 
where $n_0 \in \mathbb{Z}$ and $f_{\Delta}[n]=\big|\mathsf{f}_{i}[n]-\mathsf{f}_{i'}[n]\big|$. It means that an optimal $L$ exists if we perfectly obtain the IF information of signal modes. Nevertheless, this way of determining the optimal window length does not work in practice because: \textbf{i)} The mode's initial IF estimation $\hat{f}_{i}[n]$ in \textit{Algorithm 1} cannot be sufficiently accurate especially for large $f_{\Delta}[n]$; \textbf{ii)} Fig. \ref{fig4} only represents the two-mode case when $I=2$, and there actually exist $\mathrm{C}_I^2$ sub-optimal window lengths for more general cases when $I > 2$. It will become difficult to weigh an optimal value of $L$ from multiple sub-optimal values according to (\ref{TSP_eq28}).

Alternatively, an optimal window length can be determined by assessing the trade-off between signal bias and noise variance. It is seen from Fig. \ref{fig4} that the window length significantly affects the accuracy of IF enhancement. On the one hand, the bias for IF estimation needs to be controlled by using a limited window length. On the other hand, a large window length is expected to reduce the noise variance as well as the  separation interference especially in low SNR situations. 

Note that our previous analysis assumes that the IF of each mode, $\mathsf{f}[n]$, is known and kept at a constant value  within the analyzed window. However, more practical signals with varying IFs and estimated IF errors should be taken into account for further theoretical analysis. Let us consider a signal $x[n]$ whose IF is approximately regarded as a linear frequency-modulated (LFM) segment within the range $[-L/2,L/2]$, as any nonstationary signal can be viewed as the combination of a group of LFM segments. Defining an LFM signal segment $x[n]=\cos\big(2\pi T\sum_{m=0}^n {f}[m]\big)$ with the IF $f[n]=\mathsf{f}_0 + \mathsf{r}_0 n T$, where $\mathsf{f}_0$ and $\mathsf{r}_0$ denote initial frequency and chirp rate of the LFM segment, respectively. Without specially referring to any particular IF estimator, let us assume the IF estimation error, $\Delta[n]= \hat{f}[n]-f[n]$, which is modeled as a zero-mean Gaussian noise process with a variance $\sigma_e^2$.

The signal's nonstationarity and inaccurate IF estimation will lead to a deterministic bias for IF enhancement, thus we  derive the bias expression of the KPA method by considering time-varying IF with estimation error. According to (\ref{TSP_eq10}), the encoded LFM signal segment becomes:
\begin{align} \label{TSP_eq29}
{z_x}[n] =& \exp\bigg\{j\frac{\sin(2\pi \mathsf{f}_0 nT+\pi \mathsf{r}_0 n^2T^2)}{\mathsf{f}_0+\mathsf{r}_0 nT}\bigg\}.
\end{align}
The enhanced signal using the KPA method is derived as in (\ref{TSP_eq30}), which is presented on the top of the current page, 
where $a_1=0$, $a_2=\nint{\frac{f_s}{4\hat{f}[n]}}$, $b_1=\hat{f}[n] \sin\big(2\pi \hat{f}[n] lT\big)/2$, $b_2=\hat{f}[n]
\cos\big(2\pi \hat{f}[n] lT\big)/2$, and
$\hat{f}[n]=\mathsf{f}_0+\mathsf{r}_0 nT+\Delta[n]$.

To obtain simplified bias expressions from (\ref{TSP_eq30}), we next consider the bias analysis in two different cases: constant $f[n]$ with the impact of estimated IF error as well as  time-varying $f[n]$ with perfect IF estimation. The mathematical bias expressions for these two cases are introduced as follows.

\subsubsection{Bias analysis for constant $f[n]$ with $\Delta[n]\neq 0$}
\setcounter{equation}{30}
In this case, the estimated IF is $\hat{f}[n]=\mathsf{f}_0+\Delta[n]$, i.e., $\mathsf{r}_0=0$. 
The bias of the KPA method is thus derived from (\ref{TSP_eq30}):
\begin{align} \label{TSP_eq31}
\mathcal{B}_1[n]= &x[n]-\mathbb{E}\big\{\hat{x}[n] \big\}\\=&x[n]- x[n]\cdot \mathbb{E}\Bigg\{ \frac{\hat{f}[n]}{\mathsf{f}_0}\mathsf{E}_{l}\Big\{\sin\big(2\pi \hat{f}[n] lT\big)\sin\big(2\pi \mathsf{f}_0 lT\big)
\notag\\& +\cos\big(2\pi \hat{f}[n] lT\big)\sin\big(2\pi \mathsf{f}_0 T(l+\nint{\frac{f_s}{4\hat{f}[n]}})\big)\Big\}\Bigg\}, \notag
\end{align}
where $\mathsf{E}_{l}$ denotes the mean computation over $l$. When the error variance $\sigma_e^2$ is small, $\mathcal{B}_1[n]$ approximates to:
\begin{align} \label{TSP_eq32}
\mathcal{B}_1[n]  \approx
x[n]\cdot \mathbb{E}\bigg\{1-\frac{\hat{f}[n]}{\mathsf{f}_0}\mathsf{E}_{l}\Big\{\cos\big(2\pi \Delta[n] lT\big)\Big\}\bigg\},
\end{align}
which shows that the value of the bias depends on $\Delta$, $L$ and  $x[n]$. The maximum deviation occurs at a peak or a
valley of the sinusoidal waveform, whereas the deviation is zero when $x[n]=0$, meaning that the effect of the inaccurate IF estimation is to reduce the signal amplitude and slightly affect the signal frequency. This indicates the KPA method for the purpose of spectral restoration can tolerate certain amount of IF estimate error.

\subsubsection{Bias analysis for time-varying $f[n]$ with $\Delta[n]=0$}

\begin{figure}[!t]
\vspace{-3mm}
\hspace{-4mm}
\includegraphics[width=3.8 in]{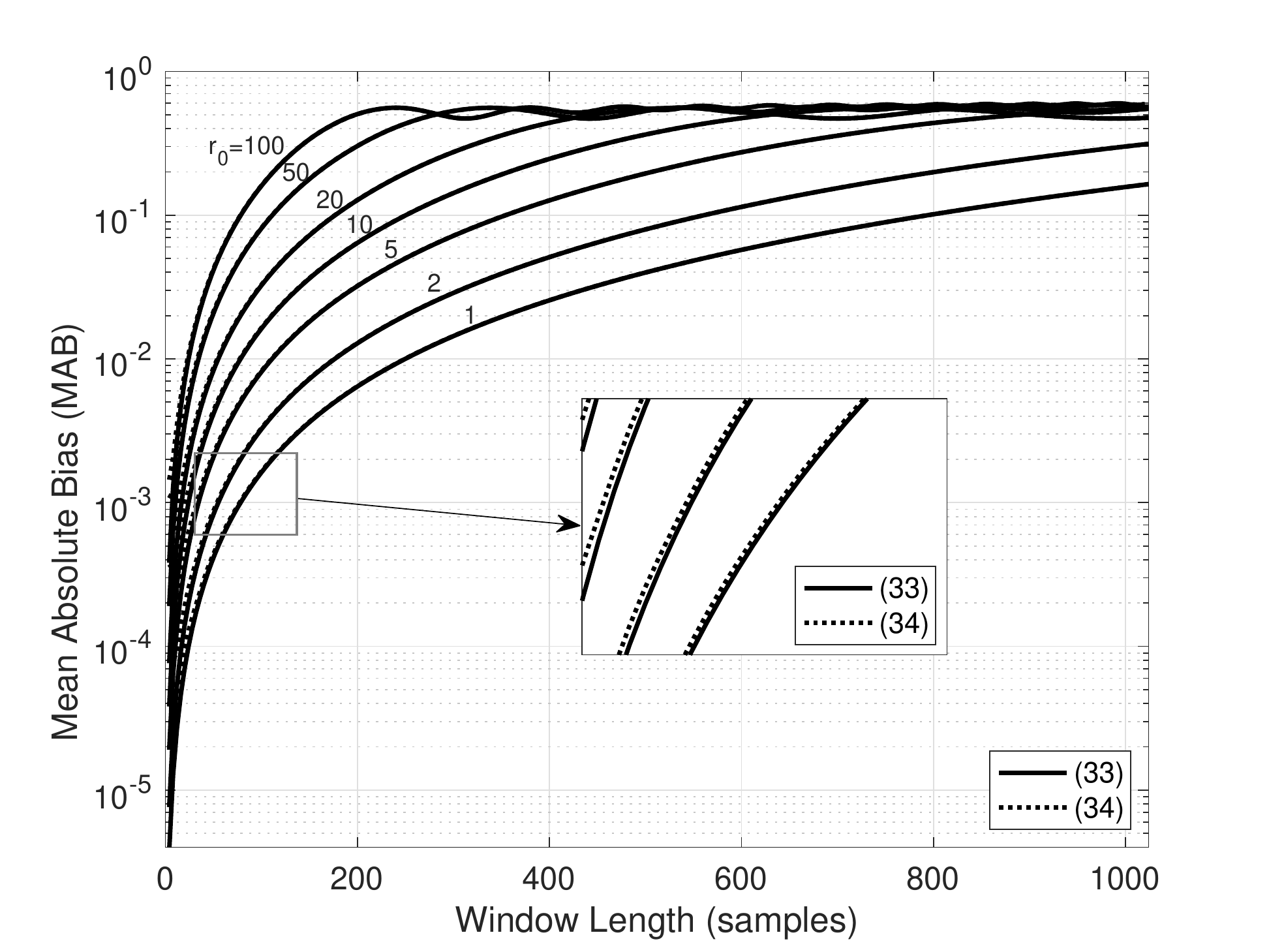}
\caption{Mean absolute biases (MABs) computed from (\ref{TSP_eq33}) and (\ref{TSP_eq34}) versus different window length $L$ with $\rho=1$, $\mathsf{f}_0$ = 50 Hz, $f_s$ = 1024 Hz, $N=1024$, and $\mathsf{r}_0\in\{1,2,5,10,20,50,100\}$.}
\label{fig4b}
\end{figure}

In the similar manner as in (\ref{TSP_eq31}), the bias expression for time-varying $f[n]$ with $\mathsf{r}_0\neq 0$ is simplified into:
\begin{align}\label{TSP_eq33}
\mathcal{B}_2[n]=& x[n]-\mathsf{E}_{l}\Bigg\{\mathcal{D}_1\sin^2\big(2\pi lT(\mathsf{f}_0 +\mathsf{r}_0 nT)\big) \\ &
\cdot \cos\big(2\pi \mathsf{f}_0 nT+\pi  \mathsf{r}_0 n^2T^2 +\pi   \mathsf{r}_0 l^2T^2\big)\notag\\ & + \mathcal{D}_2\cos^2\Big(2\pi  lT(\mathsf{f}_0 +\mathsf{r}_0 nT)\Big)\cos\Big(2\pi \mathsf{f}_0 nT\notag\\
&+\pi  \mathsf{r}_0 n^2 T^2  +\pi  \mathsf{r}_0 T^2\big(l+\nint{\frac{f_s}{4(\mathsf{f}_0 +\mathsf{r}_0 nT)}}\big)^2\Big)\Bigg\},\notag
\end{align}
where
\begin{align}
\begin{cases}
\mathcal{D}_1=\frac{ f^2[n]}{\big(f[n]+\mathsf{r}_0 lT\big)\big(f[n]-\mathsf{r}_0 lT\big)};\\
\mathcal{D}_2= \frac{ f^2[n]}{\big(f[n]+\mathsf{r}_0 lT+\mathsf{r}_0\frac{f_s}{4f[n]}\big)\big(f[n]-\mathsf{r}_0 lT-\mathsf{r}_0 \frac{f_s}{4f[n]}\big)}.\notag
\end{cases}\notag
\end{align}
When the value of  $\frac{f_s}{4f[n]}$ is small, the bias expression in (\ref{TSP_eq33}) degrades into:
\begin{align} \label{TSP_eq34}
 \mathcal{B}_2[n]  \approx& \cos(2\pi \mathsf{f}_0 nT+\pi  \mathsf{r}_0 n^2T^2) \\
 &-\mathsf{E}_{l}\bigg\{\mathcal{D}_1 \cos\big(2\pi \mathsf{f}_0 nT+\pi \mathsf{r}_0  n^2T^2 +\pi \mathsf{r}_0 l^2T^2\big)\bigg\}, \notag
\end{align}
which equals to zero when $\mathsf{r}_0=0$. The expressions in (\ref{TSP_eq32}) and (\ref{TSP_eq34})  show that the most significant impact for IF enhancement is the chirp rate $\mathsf{r}_0$ in $\mathcal{B}_2[n]$ rather than IF estimation error $\Delta$ in $\mathcal{B}_1[n]$. Fig. \ref{fig4b} presents the mean absolute biases (MABs) calculated by (\ref{TSP_eq33}) and (\ref{TSP_eq34})
as the window length $L$ increases with varying $\mathsf{r}_0\in\{1,2,5,10,20,50,100\}$. It is seen that the approximate expression (\ref{TSP_eq34}) well fits the exact expression (\ref{TSP_eq33}) under the condition $\mathsf{f}_0$ = 50 Hz  and $f_s$ = 1024 Hz.

Since the KPA method is more concerned with the effect of bias on frequency information rather than amplitude information, the optimal window size for IF enhancement  can be chosen by limiting the bias quantity in (\ref{TSP_eq34}).  Specifically, the deterministic bias due to the signal's nonstationarity can be minimized by constraining the maximum of the elementary part $\pi \mathsf{r}_0 l^2T^2$ in (\ref{TSP_eq34}): 
\begin{align} \label{TSP_eq35}
 \max_{l}\bigg\{\bigg|\frac{\pi \mathsf{r}_0 l^2}{f_s^2} \bigg|\bigg\}
< \varepsilon_2,~~~ l\in \Big[-\frac{L}{2},\frac{L}{2}\Big]
\end{align}
where $\varepsilon_2$ is a constant coefficient used to control the quantity of MAB. As a result, the criterion of choosing the window length $L$ is derived as a function of  $\mathsf{r}_0$, $\varepsilon_2$, and $f_s$:
\begin{align} \label{TSP_eq36}
 L \leq 2f_s \sqrt{\frac{\varepsilon_2}{\pi \mathsf{r}_0}},
\end{align}
which implies that the window size should be reduced when a strict MAB (a small value of $\varepsilon_2$) is required. Furthermore, as the value of $\mathsf{r}_0$ decreases, large window sizes are allowed. This is reasonable because no window or an infinite window length can be used when $\mathsf{r}_0=0$. Consequently, the window length $L$ in a practical application is selected according to the practical MAB requirement. For instance, an optimal value of $L$ = 36 is selected if a strict bias $\varepsilon_2<10^{-2}$ is required when $\mathsf{r}_0$ = 10 and $f_s$ = 1024 Hz.  In practice, the chirp rate $\mathsf{r}_0$ can be coarsely estimated from the initial IF estimation in \textit{Algorithm 1}. Note that a coarse estimation of $\mathsf{r}_0$ is sufficient because the window size changes slowly with the variation of $\mathsf{r}_0$, as shown in Fig. \ref{fig4b}. Besides, the window length can be enlarged by increasing the sampling frequency $f_s$.

\textit{Remark.}  The above biases have been analyzed based on LFM signal model in (\ref{TSP_eq29}), but the derived theoretical result in (\ref{TSP_eq36}) is also suitable for other nonstationary signals with arbitrary frequency variation. The reason lies in that every segment of a nonstationary signal within a limited lag window can be approximately viewed as an LFM waveform according to the Weierstrass approximation theorem.

\subsection{Complexity Analysis of the Proposed ETFR-MD}

The overall computational complexity of the proposed ETFR-MD should include the initial IF estimation in \textit{Algorithm 1}, the IF enhancement using the KPA method, and the IA extraction via STFT. The main merit of the KPA method for IF enhancement is that the enhanced signal is obtained by performing the phase averaging operation in (\ref{TSP_eq16}) instead of detecting time-slice peaks of STFD in (\ref{TSP_eq4}). From the viewpoint of implementation, the operation in (\ref{TSP_eq16}) is much less complex compared to that in (\ref{TSP_eq4}) by avoiding DFT computation and peak search process. Therefore, the mathematical description of the computational complexity for the KPA method is  
$\mathcal{O}(K I N)$, where $K$ is the number of iterations, $I$ is the number of signal modes, and $N$ denotes the number of signal samples. The proposed KPA method not only has a linear  complexity, but also is compatible with  parallel computation. In addition, the IAs of signal modes are recovered through the computation of STFT in (\ref{TSP_eq18}), which has a complexity $\mathcal{O}(N\log{}N)$.

Despite the low complexity of IF and IA recovery in the ETFR-MD, its practical implementation requires an initial IF estimation by carrying out state-of-the-art methods or the proposed method in \textit{Algorithm 1}, which implies an additional increase in the computational cost. Nevertheless, it does not significantly increase the overall complexity. Experimental results in the next section will demonstrate that the gain of enhanced representation and decomposition performance well offsets the additional  complexity on initial IF estimation.

\section{Experimental Evaluation}

\begin{figure*}[!t]
\hspace{-9mm}
\includegraphics[width=7.8 in]{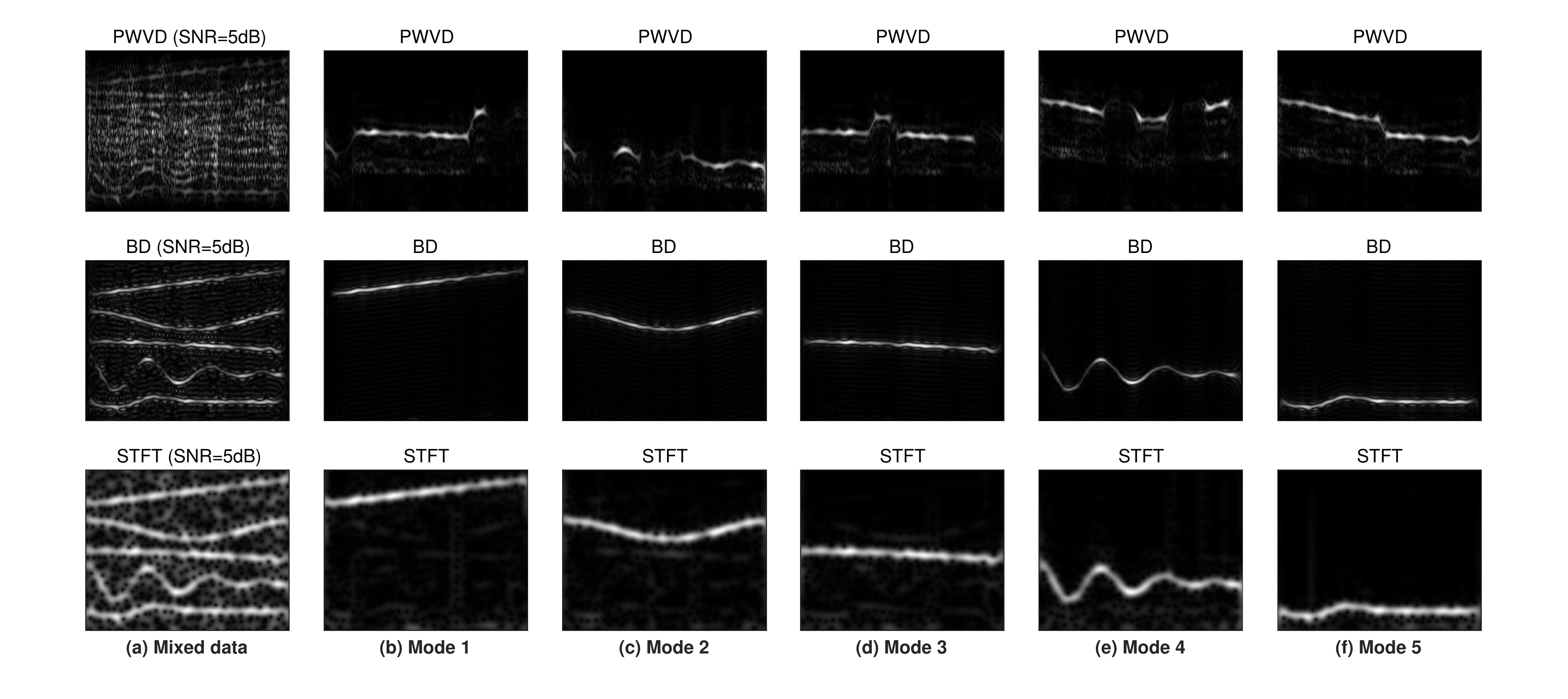}
\vspace{-11mm}
\caption{TFRs of the multi-mode signal with five closely-spaced FM modes  at SNR = 5 dB: \textbf{(a)} The first column presents the TFRs of the noisy multi-mode signal computed  by PWVD, BD, and STFT, respectively. \textbf{(b)-(f)} The rest of columns present the TFRs of the five separated FM modes using the proposed KPA method, where the initial IF estimation is implemented based on PWVD, BD, and STFT, respectively.}
\label{fig5}
\end{figure*}

\begin{figure*}[!t]
\centering
\includegraphics[width=6.5 in]{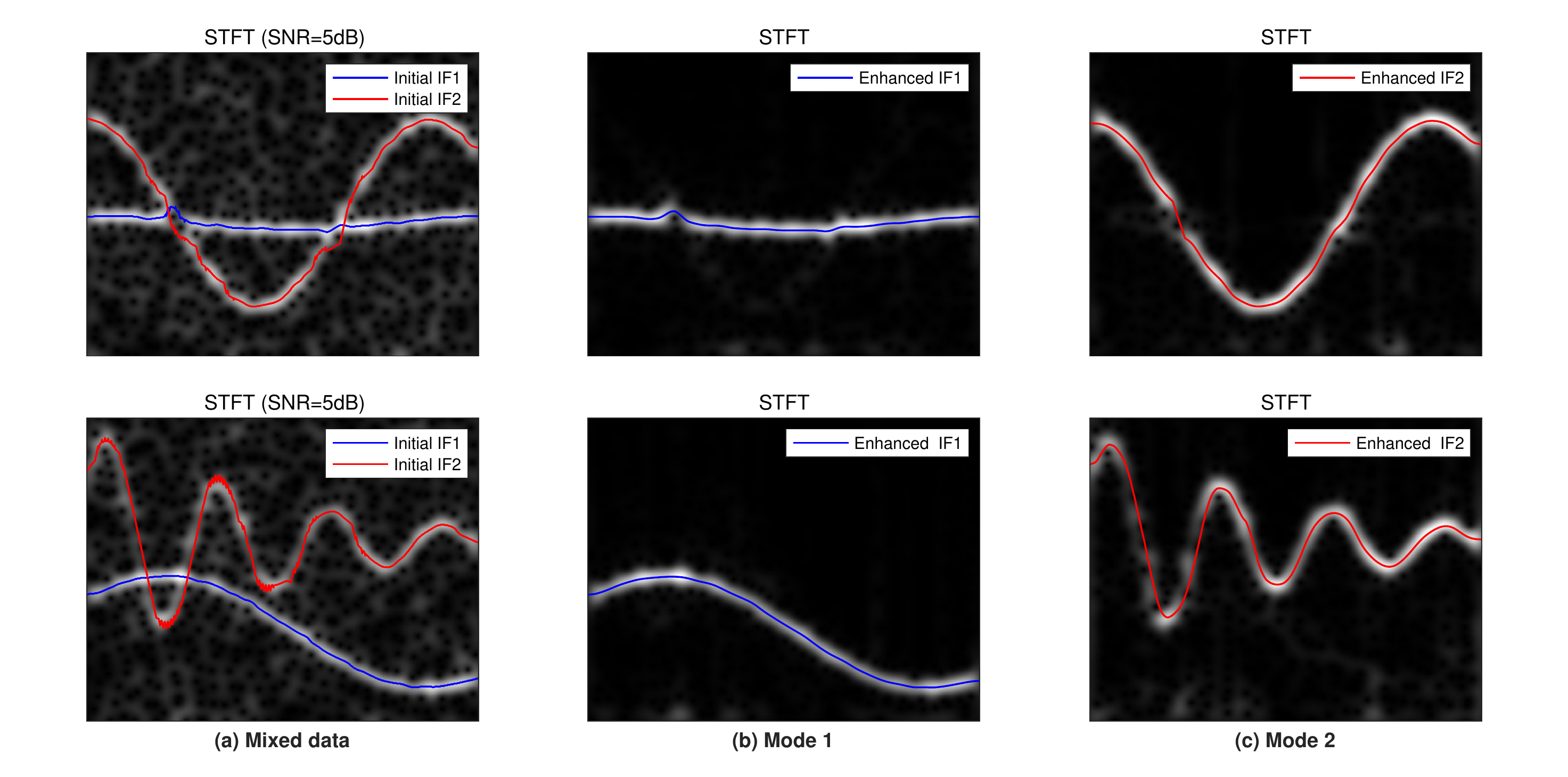}
\vspace{-5mm}
\caption{STFTs of two multi-mode signals with spectrally-overlapped SFM modes at SNR = 5 dB: \textbf{(a)} The first column presents the STFTs of the two different noisy multi-mode signals, where the red and blue curves represent the initial IF estimation of each mode. \textbf{(b)-(c)} The rest of columns present the STFTs of the two separated SFM modes using the KPA method, where the red and blue curves represent the enhanced IF estimation of each mode.}
\label{fig6}
\end{figure*}

In this section, we demonstrate the benefits of the proposed ETFR-MD method in dealing with multi-mode signals with close or crossing IFs in terms of IF enhancement, enhanced TFR, and mode decomposition:
\begin{itemize}
\item \textit{IF enhancement}: Experimental results of the proposed KPA method are firstly given to illustrate its effectiveness in enhancing initial IF estimates.

\item \textit{Enhanced TFR}: The enhanced time-frequency representations by combining  extracted IF and IA information are compared with the  SST \cite{Auger6633061} and SET \cite{yu2017synchroextracting} methods.

\item \textit{Mode decomposition}:  The proposed ETFR-MD for mode decomposition is  evaluated via a comparison against SST \cite{Auger6633061}, SST4 \cite{Pham7885114}, MSST \cite{Yu8458385}, and SET \cite{yu2017synchroextracting}, which are state-of-the-art  methods in TF representation and mode reconstruction. The decomposed modes are assessed via mean square error (MSE) and Output SNR metrics \cite{Auger6633061}.
\end{itemize}

In our study, the three typical multi-mode FM signals  with close or crossing IFs in Fig. \ref{fig1} are selected for the computer simulation.  The initial IF estimation of these multi-mode signals is implemented using \textit{Algorithm 1}. Note that an alternative IF estimation method, e.g., ridge detection related ones, can be used when non-overlapping FM modes are involved. Since the number and type of signal modes are unknown, we set empirical parameters for the entire multi-mode signal, thus forcing a compromise in the parameter setting for signals with different modes. Specifically, the experimental parameters in \textit{Algorithm 1} are empirically selected: $c_1=50$, $c_2=5000$, $\Delta_1=3$, $\Delta_2=0.3$, $\delta=30$. The number of iterations in the KPA method is set to $K=3$, moreover, we suggest the window length $L=32$, which is chosen so that the bias of the enhanced modes in (\ref{TSP_eq36}) is approximately equal to the magnitude of $10^{-2}$. It is assumed that the noise is additive white Gaussian noise (AWGN), and we also assume that all the signal modes exist throughout the time domain.

\begin{figure*}[!t]
\hspace{-2mm}
\includegraphics[width=7.2 in]{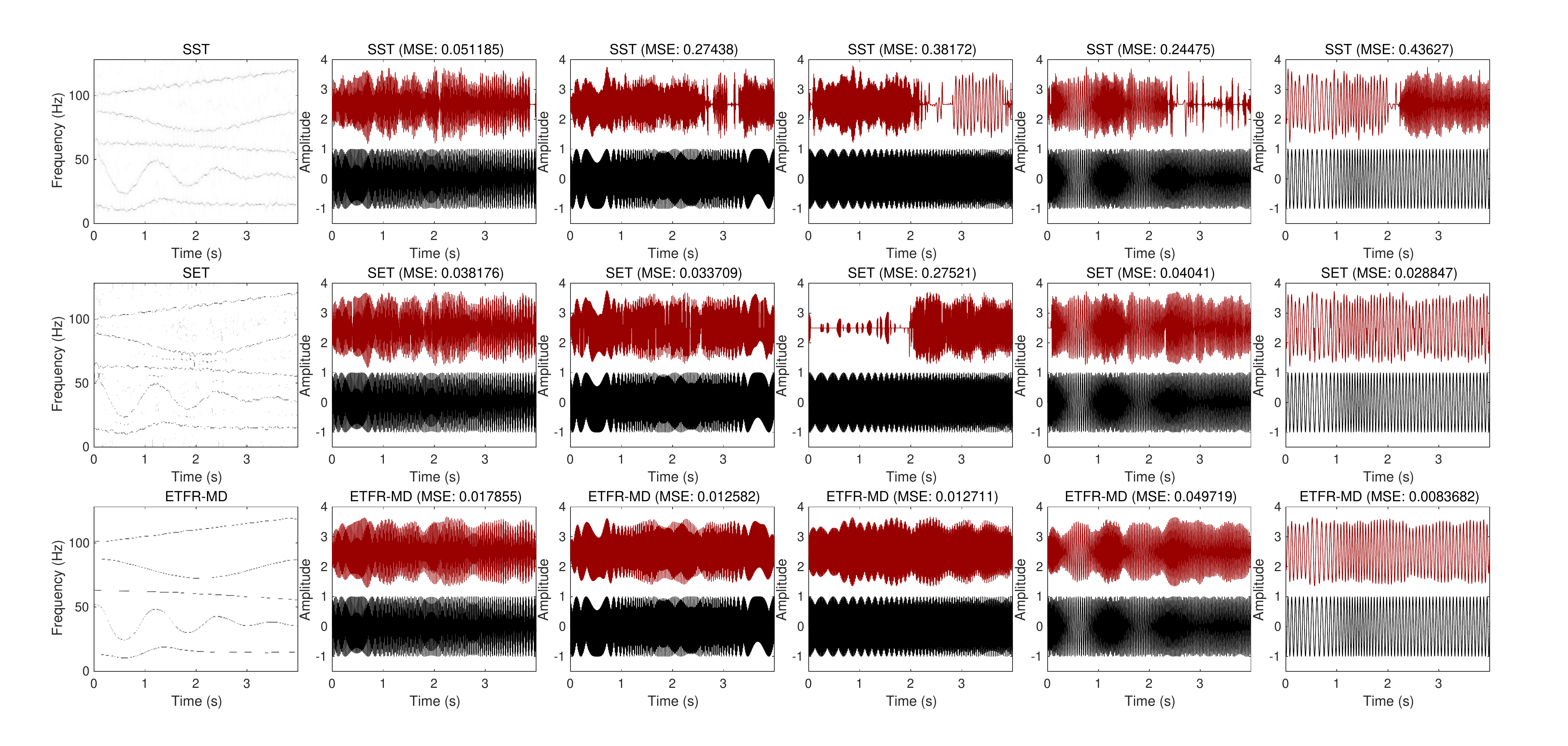}
\vspace{-9mm}
\caption{Mode decomposition of the multi-mode signal with five closely-spaced FM modes using three different TF decomposition methods at SNR = 10 dB. \textbf{(First Row) } SST with $d=15$ \cite{Auger6633061} and its five decomposed FM modes. \textbf{(Second Row)} SET \cite{yu2017synchroextracting} and its five decomposed FM modes. \textbf{(Third Row)} The proposed ETFR-MD and its five decomposed FM modes. The MSE value is annotated on the top of each decomposed FM mode, and a bias of 2.5 is added to the estimated FM modes (in red) for better visual quality comparison with the true FM modes (in black).}
\label{fig8}
\end{figure*}

\begin{figure*}[!t]
\centering
\hspace{-3mm}
\subfigure[Weakly-modulated FM modes.]{
\includegraphics[width=3.5 in]{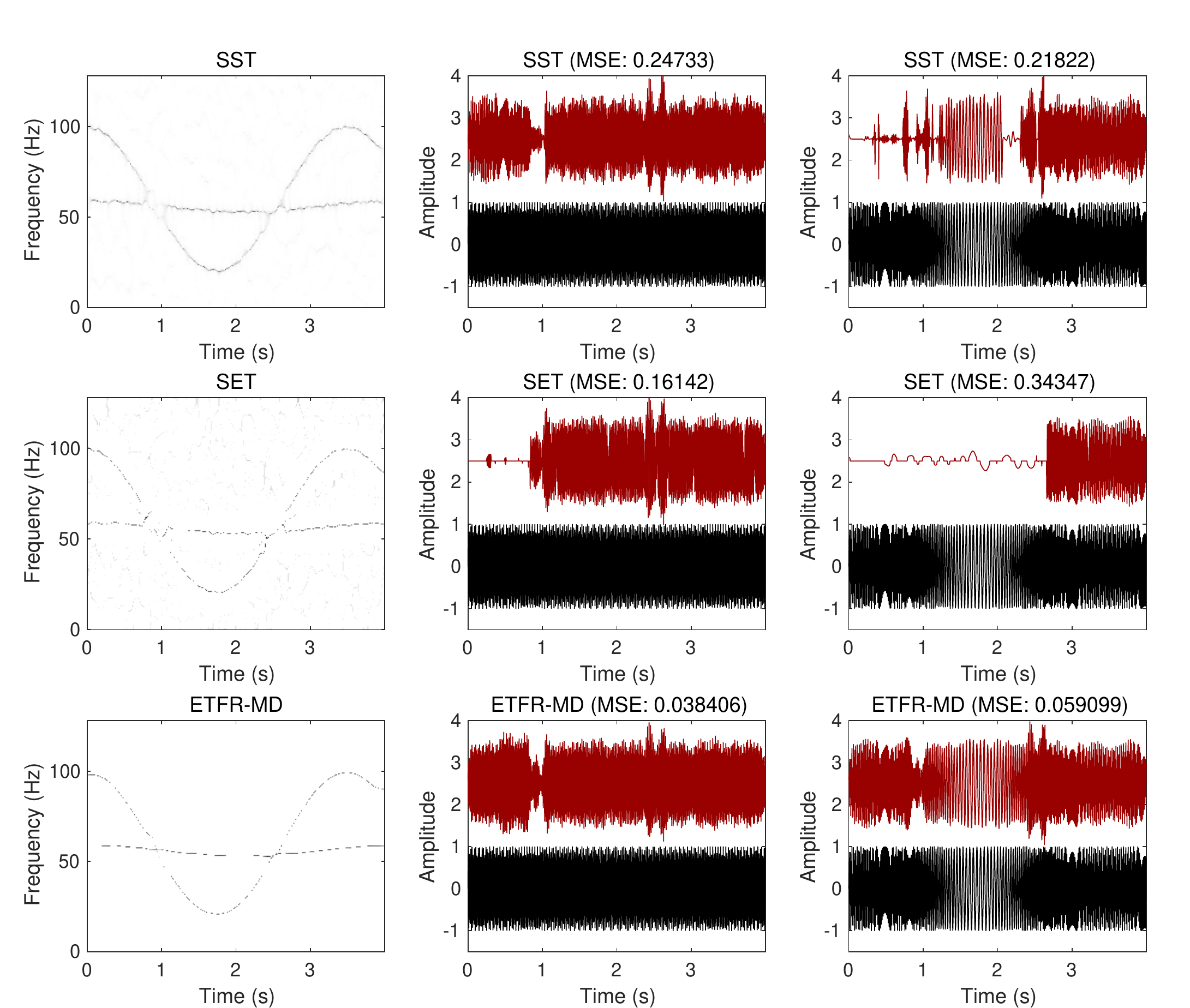}
}
\hspace{-3mm}
\subfigure[Strongly-modulated FM modes.]{
\includegraphics[width=3.5 in]{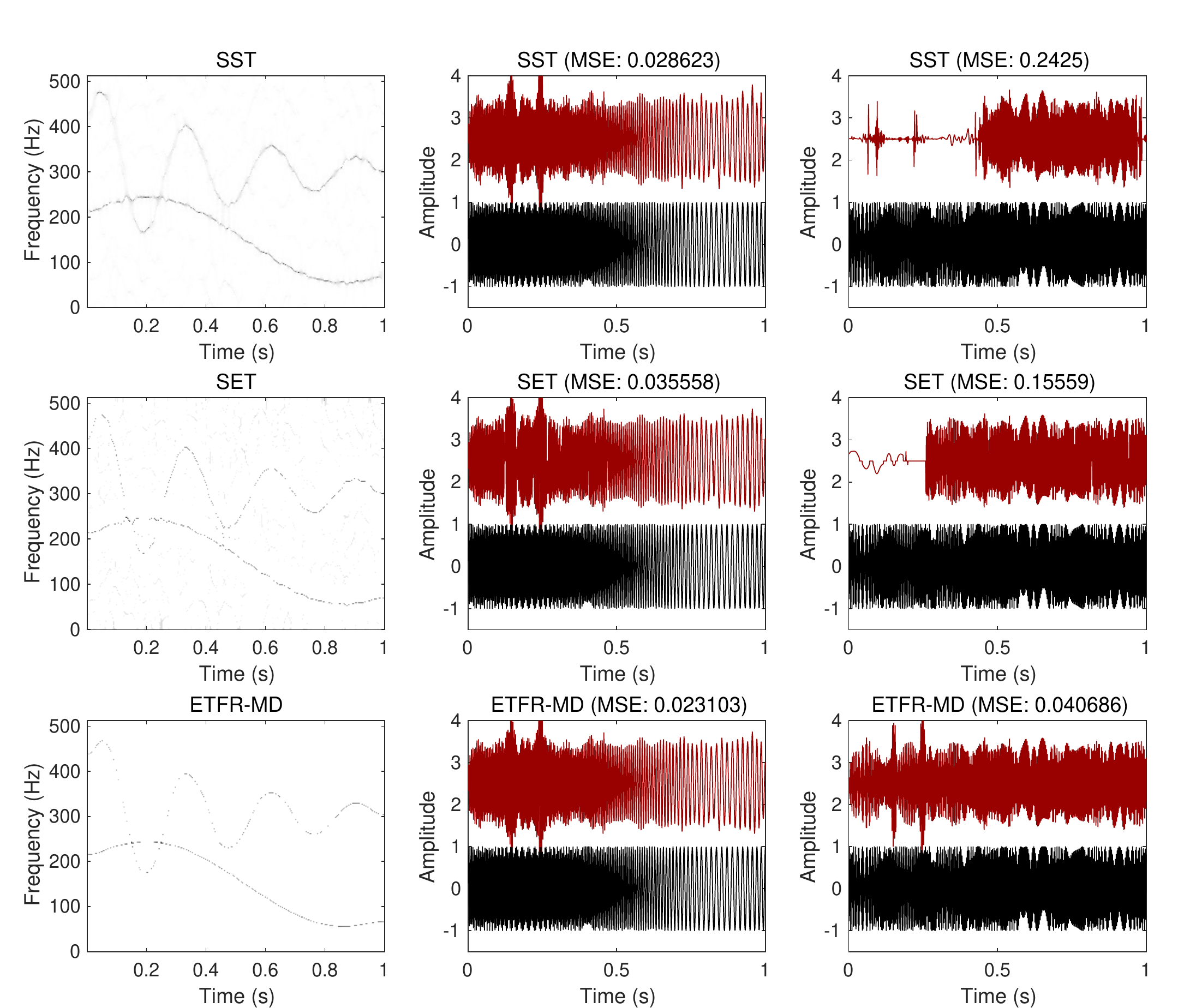}
}
\caption{Mode decomposition of two multi-mode signals with crossing FM modes using three different TF decomposition methods at SNR = 10 dB. Rows from up to down correspond to SST, SET, and the proposed ETFR-MD, respectively. \textbf{(a)} Multi-mode signal with weakly-modulated FM modes. \textbf{(b)} Multi-mode signal with strongly-modulated FM modes. The MSE value is annotated on the top of each decomposed FM mode, and a bias of 2.5 is added to the estimated FM modes (in red) for better visual quality comparison with the true FM modes (in black).}
\label{fig9}
\end{figure*}

\begin{figure*}[!t]
\hspace{-4mm}
\subfigure[Five closely-spaced FM modes.]{
\includegraphics[width=2.5 in]{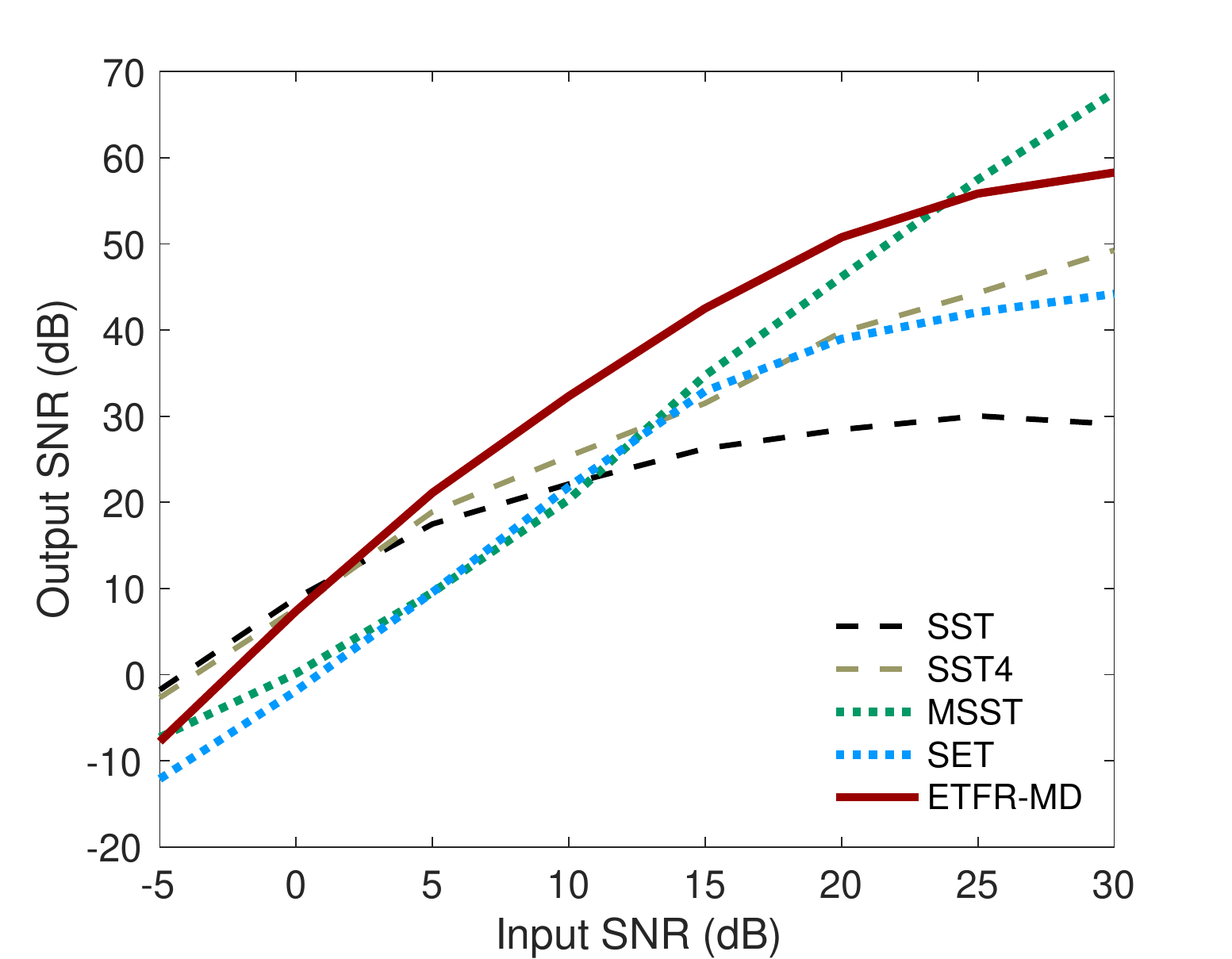}
}
\hspace{-6mm}
\subfigure[Two crossing weakly-modulated FM modes.]{
\includegraphics[width=2.5 in]{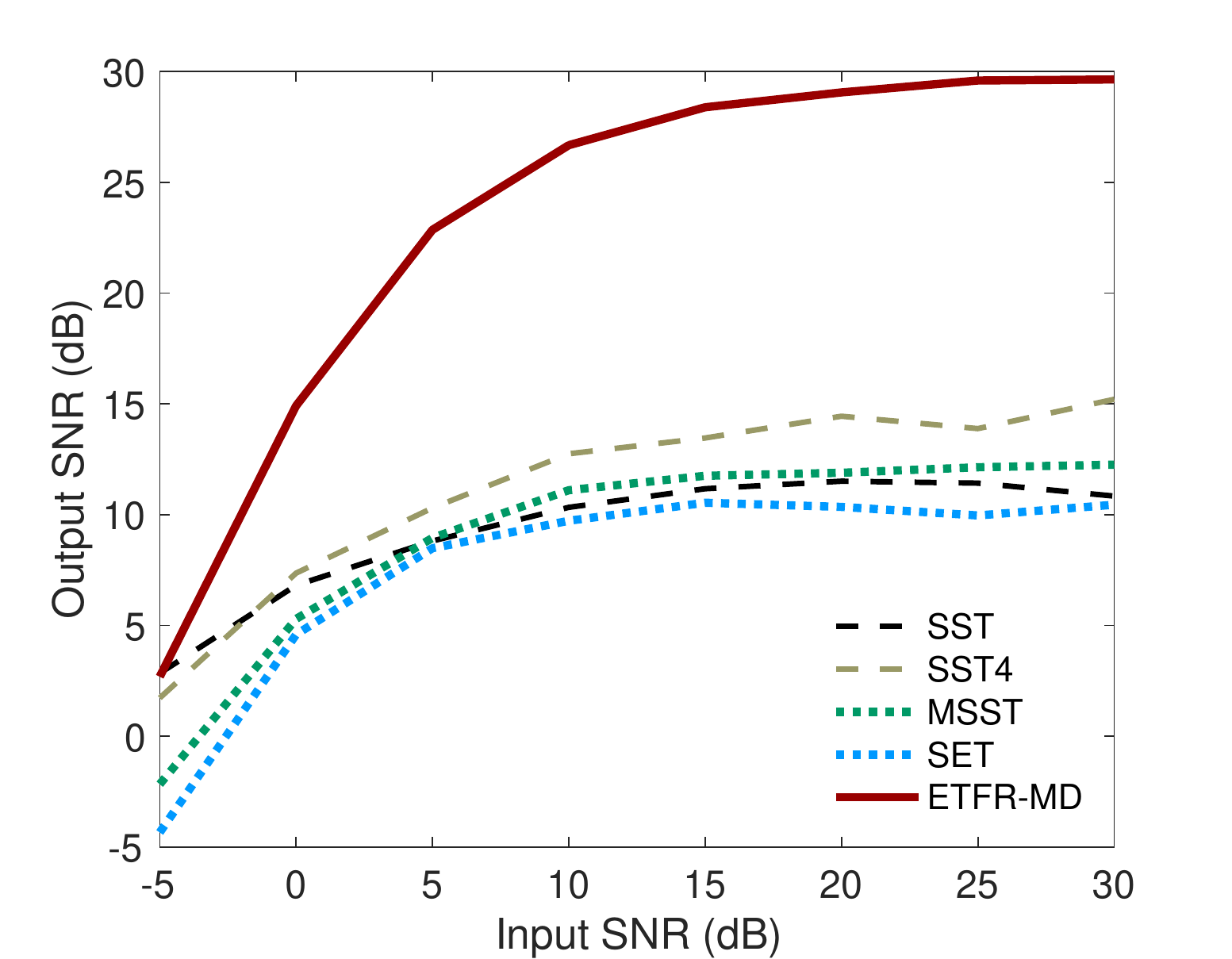}
}
\hspace{-6mm}
\subfigure[Two crossing highly-modulated FM modes.]{
\includegraphics[width=2.5 in]{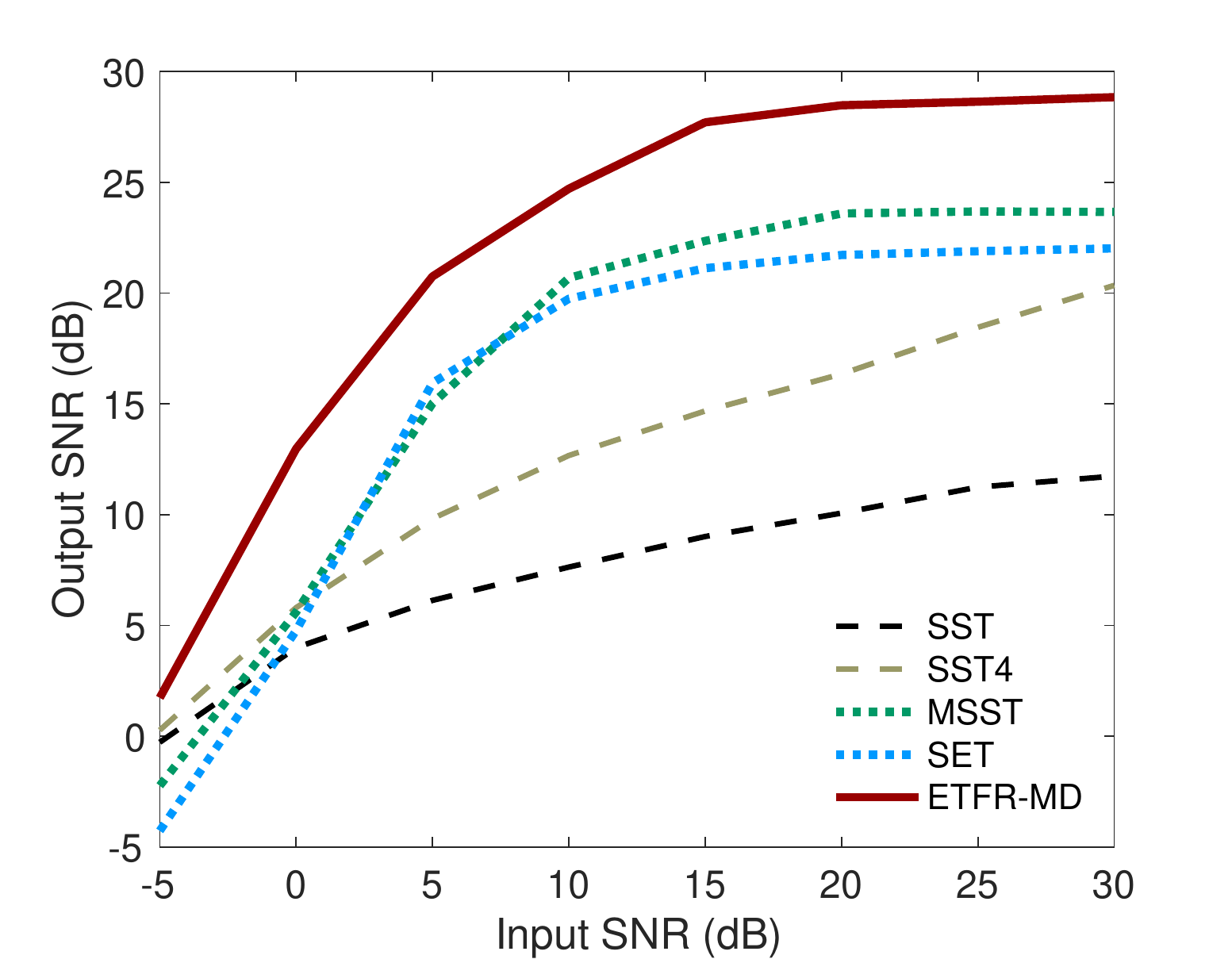}
}
\caption{Output SNRs of reconstructed signal modes versus different input SNR levels (50 trials are implemented at each SNR) using SST with $d=10$ \cite{Auger6633061}, SST4 with $d=10$ \cite{Pham7885114}, MSST with $d=10$ and 6 iterations \cite{Yu8458385},  SET \cite{yu2017synchroextracting},  and the proposed ETFR-MD: \textbf{(a)} Five closely-spaced FM modes. \textbf{(b)} Two spectrally-crossing and weakly-modulated SFM modes. \textbf{(c)} Two spectrally-crossing and  strongly-modulated SFM modes.}
\label{fig10}
\end{figure*}

\subsection{Enhanced IF Estimation}

Firstly, we consider the multi-mode signal whose five IF laws are non-overlapping but spatially close in TF domain. As shown in the first column of Fig. \ref{fig5}, two quadratic TFRs, i.e., PWVD and B-distribution (BD) \cite{950779}, are also simulated for comparison with STFT at SNR = 5 dB. The remaining columns of Fig. \ref{fig5} show the corresponding TFRs of the five enhanced modes using the KPA method in (\ref{TSP_eq15}). It is worth noting that the PWVD generating strong CTs significantly impacts the initial IF estimation, and fails to implement the KPA method. In contrast, the BD has both high TF resolution and maximum CT reduction, thus obtaining accurate IF estimate. Although BD exhibits more concentrated TFR than STFT, we can still implement the initial IF estimation from the simpler STFT. It is obvious from the second and third rows of Fig. \ref{fig5} that the IF ridge of each mode  can be clearly recognized, based on which the IF enhancement of each mode becomes possible. 

Secondly, to test initial IF estimation and   IF enhancement using the KPA method, two multi-mode signals with intersected IF laws are considered. It is seen from the first column of Fig. \ref{fig6} that the initial IFs based on STFT could be well detected despite the interference of crossing TF points, and then the IF of each mode is further enhanced via (\ref{TSP_eq17}). As illustrated in the second and third columns of Fig. \ref{fig6}, the enhanced IFs are more smooth and accurate compared with initial ones. The comparative results show the superiority of using the KPA method on IF enhancement even for noisy signals at a low SNR level.

\subsection{Enhanced Time-Frequency Representation}

Next, we present the enhanced TFRs by a comparative study with the classical SST and currently reported SET to verify the robustness of the proposed ETFR-MD in (\ref{TSP_eq20}). The first columns of Fig. \ref{fig8} and Fig. \ref{fig9} show the TFR behaviors of different methods on the three multi-mode signals when SNR = 10 dB. It is observed that the proposed ETFR-MD provides the clearest and most concentrated TFRs, while somewhat more expensive since initial  IF estimation is required. For the methods of comparison, their parameters are adjusted to result in the most visually appealing TFRs, but these TFRs exhibit   frequency and amplitude variations under low SNR conditions and complex situations with close or crossing IFs.

\subsection{Time-Frequency Mode Decomposition}

Finally, we demonstrate the performance improvement of the proposed ETFR-MD method that contributes to mode decomposition. Fig. \ref{fig8} and \ref{fig9} respectively show the mode decomposition results of the above three multi-mode signals using the SST \cite{Auger6633061}, the SET \cite{yu2017synchroextracting}, and our ETFR-MD in (\ref{TSP_eq21}) when SNR = 10 dB. The MSEs of estimated signal modes calculated with ideal modes are indicated on the top of each figure. Note that both the SST and SET based mode decomposition requires an initial IF estimation of each mode, which is realized by the ridge detection method  from SST and SET, respectively. Experimental results in Fig. \ref{fig8} and \ref{fig9}  reveal the benefit of the ETFR-MD using our proposed \textit{Algorithm 1} for multi-mode FM signals with close and crossing IFs.

Considering that more quantitative measures would usefully supplement and extend the previous qualitative analysis.
In addition to SST and SET, we also compare our ETFR-MD with the SST4  \cite{Pham7885114} and the MSST \cite{Yu8458385}, two of the state-of-the-art methods, in terms of the Output SNR metric. For SST-related methods, a larger parameter $d$ means  more accurate mode reconstruction, thus $d=10$ is selected for a fair comparison. Fig. \ref{fig10} presents the Output SNRs of reconstructed signal modes versus different input SNRs using four existing methods and our ETFR-MD. It is observed from Fig. \ref{fig10} (a) that the ETFR-MD outperforms other methods in most cases for the non-overlapping multi-mode signal. While the MSST method achieves the best performance at high SNRs, it is conducted in an iterative process (6 iterations in our experiments). Fig. \ref{fig10} (b) and (c)  show that our ETFR-MD has attained much larger performance gain for spectrally-overlapping multi-mode signals, which largely benefits from  robust IF estimation and enhancement of the KPA method.

\section{Conclusion}

To address TF representation and mode reconstruction of complicated FM signals with close or crossing IFs, this paper proposes an STFT-based ETFR-MD method considering the widespread use of STFT and its invertibility. Moreover, an efficient  method is designed for initial IF estimation of multi-mode signals. The essence of the proposed ETFR-MD method lies in the separate IF and IA reconstruction of each mode, which are respectively accomplished by the KPA method and STFT coefficients.   Experimental results have verified our theoretical analysis and confirmed the central importance of IF enhancement. Our method does not rely on the parametric assumption of signal's IF model, thus extending the benefits of TF representation and decomposition to more general signals of arbitrary mode combination. However, the performance of the ETFR-MD is limited by the IF estimation in STFT domain, the estimation error of which becomes more significant as the signal's nonstationarity increases. Although our IF estimation method exceeds the cost of existing ridge detection methods, the benefits of representation and mode reconstruction may often outweigh the additional cost. To alleviate the computational cost without influencing the performance, an alternative low-complexity IF estimation method can be applied instead for non-overlapping cases. The proposed ETFR-MD method is expected to empower the development of TF representation and mode decomposition technologies  in practical applications where close or crossing signal modes are involved.

\section*{Acknowledgment}
We would like to thank the researchers for kindly sharing the source codes in their works \cite{Auger6633061,Pham7885114,yu2017synchroextracting,Yu8458385}, which are publicly available. 

\ifCLASSOPTIONcaptionsoff
  \newpage
\fi

\bibliographystyle{IEEEtran}
\bibliography{hj2020}

\end{document}